\documentclass[%
 reprint,
 amsmath,amssymb,
 aps,
prd,
superscriptaddress,
nofootinbib]{revtex4-2}

\usepackage{upgreek}
\usepackage{graphicx}
\usepackage{dcolumn}
\usepackage{bm}
\usepackage{hyperref}
\usepackage{gensymb}
\usepackage{wrapfig}
\usepackage{float}
\usepackage{braket}
\usepackage{comment}
\usepackage[title]{appendix}

\begin{document}

\title{Estimating the Energy Threshold of Phonon-mediated Superconducting Qubit Detectors Operated in an Energy-Relaxation Sensing Scheme}


\date{\today}

%
%

\author{R.~Linehan}\thanks{linehan3@fnal.gov}
\affiliation{Fermi National Accelerator Laboratory, Batavia, IL 60510, USA}

\author{I.~Hernandez}
\affiliation{Department of Physics, Illinois Institute of Technology, Chicago, IL, 60616}
\affiliation{Fermi National Accelerator Laboratory, Batavia, IL 60510, USA}

\author{D.~J.~Temples}
\affiliation{Fermi National Accelerator Laboratory, Batavia, IL 60510, USA}

\author{S.~Q.~Dang}
\affiliation{Department of Physics, Cornell University, Ithaca, NY 14853, USA}
\affiliation{Fermi National Accelerator Laboratory, Batavia, IL 60510, USA}

\author{D.~Baxter}
\affiliation{Fermi National Accelerator Laboratory, Batavia, IL 60510, USA}
\affiliation{Department of Physics \(\&\) Astronomy, Northwestern University, Evanston, IL 60208, USA}

\author{L.~Hsu}
\affiliation{Fermi National Accelerator Laboratory, Batavia, IL 60510, USA}

\author{E.~Figueroa-Feliciano}
\affiliation{Department of Physics \(\&\) Astronomy, Northwestern University, Evanston, IL 60208, USA}
\affiliation{Fermi National Accelerator Laboratory, Batavia, IL 60510, USA}

\author{R.~Khatiwada}
\affiliation{Department of Physics, Illinois Institute of Technology, Chicago, IL, 60616}
\affiliation{Fermi National Accelerator Laboratory, Batavia, IL 60510, USA}


\author{K.~Anyang}
\affiliation{Department of Physics, Illinois Institute of Technology, Chicago, IL, 60616}
\affiliation{Fermi National Accelerator Laboratory, Batavia, IL 60510, USA}

\author{D.~Bowring}
\affiliation{Fermi National Accelerator Laboratory, Batavia, IL 60510, USA}

\author{G.~Bratrud}
\affiliation{Department of Physics \(\&\) Astronomy, Northwestern University, Evanston, IL 60208, USA}
\affiliation{Fermi National Accelerator Laboratory, Batavia, IL 60510, USA}

\author{G.~Cancelo}
\affiliation{Fermi National Accelerator Laboratory, Batavia, IL 60510, USA}

\author{A.~Chou}
\affiliation{Fermi National Accelerator Laboratory, Batavia, IL 60510, USA}

\author{R.~Gualtieri}
\affiliation{Department of Physics \(\&\) Astronomy, Northwestern University, Evanston, IL 60208, USA}
\affiliation{Fermi National Accelerator Laboratory, Batavia, IL 60510, USA}

\author{K.~Stifter}\thanks{Presently at SLAC National Accelerator Laboratory, Menlo Park, CA 94025, USA}
\affiliation{Fermi National Accelerator Laboratory, Batavia, IL 60510, USA}

\author{S.~Sussman}
\affiliation{Fermi National Accelerator Laboratory, Batavia, IL 60510, USA}


\begin{abstract}

In recent years, the lack of a conclusive detection of WIMP dark matter at the 10~GeV/c\(^{2}\) mass scale and above has encouraged development of low-threshold detector technology aimed at probing lighter dark matter candidates. Detectors based on Cooper-pair-breaking sensors have emerged as a promising avenue for this detection due to the low (meV-scale) energy required for breaking a Cooper pair in most superconductors. Among them, devices based on superconducting qubits are interesting candidates for sensing due to their observed sensitivity to broken Cooper pairs. We have developed an end-to-end G4CMP-based simulation framework and have used it to evaluate performance metrics of qubit-based devices operating in a gate-based ``energy relaxation'' readout scheme, akin to those used in recent studies of qubit sensitivity to ionizing radiation. We find that for this readout scheme, the qubit acts as a phonon sensor with an energy threshold ranging down to \(\simeq\)0.4~eV for near-term performance parameters. 
\end{abstract}
\maketitle

\setlength{\parskip}{0pt}



\section{Introduction}
\label{sec:Introduction}

A wealth of astrophysical and cosmological evidence points to the abundance of a massive, cold, non-baryonic ``dark'' matter (DM) in today's universe~\cite{Arbey}. Despite dark matter's abundance, experiments designed to directly observe it in the lab have so far only yielded null results~\cite{ADMX,LZ,XENON1T}, and as a result many of its fundamental properties such as its mass are not well known. Notably, direct detection experiments searching for particle-like DM candidates continue to exclude WIMP dark matter with masses in the GeV/c\(^{2}\)-TeV/c\(^{2}\) mass range, which has spurred interest in searching for other well-motivated particle-like DM candidates at lower masses down to 50 keV/c\(^{2}\)~\cite{SnowmassLowThresholdReport}. 

\begin{figure}[t!]
\centering
\includegraphics[width=\linewidth]{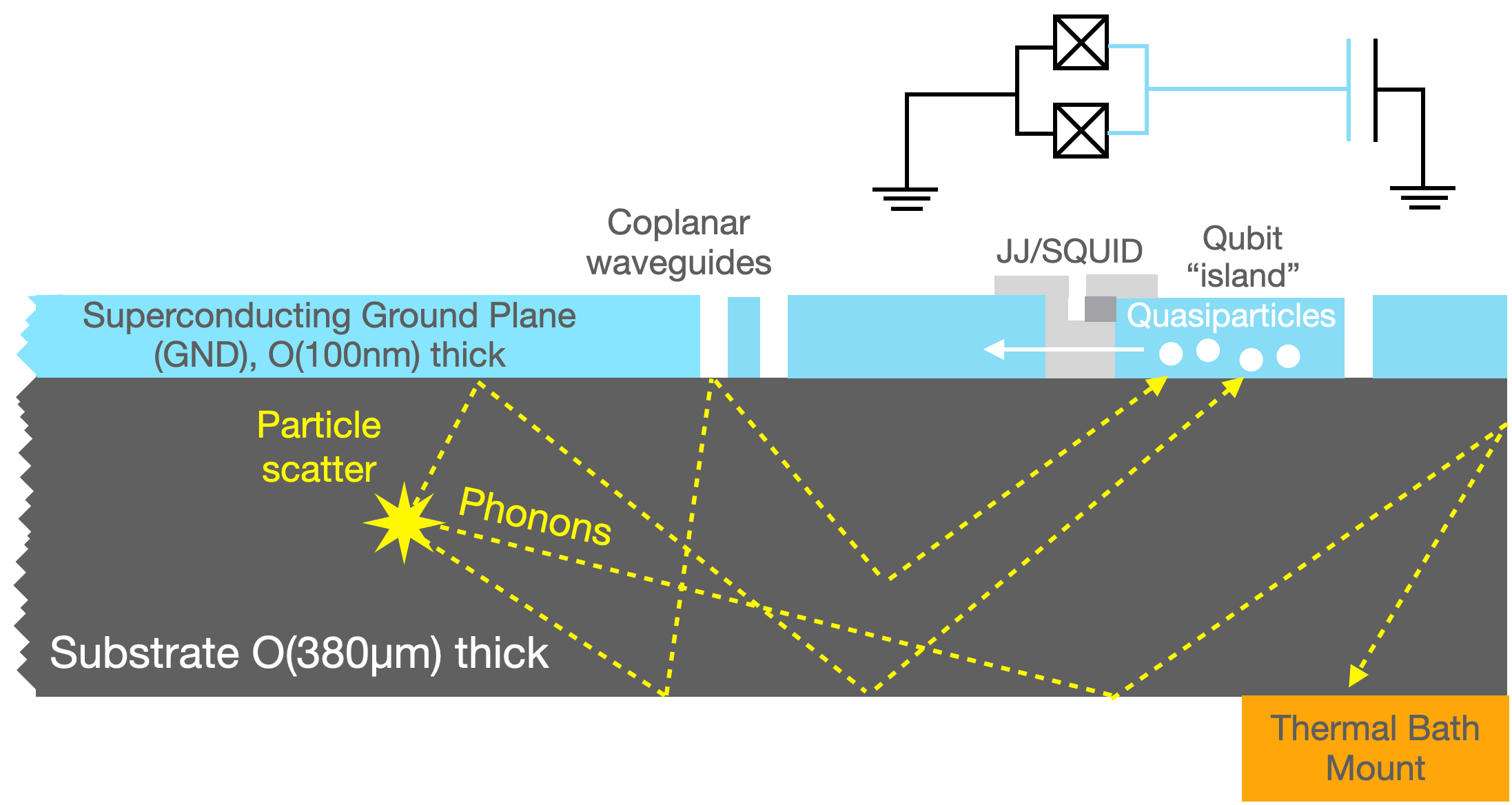}
\caption{A cross-sectional diagram of a segment of a qubit chip (not to scale), showing the substrate in gray, superconductor in blue, and thermalization mounts in orange. The qubit island and region around the JJ/SQUID are regions where enhanced QP density may be sensed. A simple circuit representation is shown above to illustrate how the qubit island (blue segment) is formed from a SQUID loop (the ``inductor'') and capacitive gap between the island and ground plane.}
\label{fig:QubitChipSideViewDiagram}
\end{figure} 

While lower-mass dark matter searches benefit from an increased flux of DM passing through the detector relative to traditional WIMP searches, this rate increase comes at the cost of a drop in the average energy deposited during a DM interaction and a corresponding drop in the detector energy threshold required for observing scatters. For example, to observe inelastic scattering of a DM particle of mass \(\simeq\)1~MeV/c\(^{2}\) on a target, a threshold of \(\simeq\)1~eV or lower is required~\cite{SnowmassLowThresholdReport}. This requirement has launched community-wide effort to develop and deploy sensors capable of observing sub-eV energy depositions. ``Pairbreaking'' sensors, which operate near 10~mK and by sensing broken Cooper pairs (Bogoliubov Quasiparticles/QPs) in superconductors, form a popular class of such low-threshold devices. Common among these are microwave kinetic inductance detectors (MKIDs) and devices based on transition-edge-sensors (TESs), both of which consist of thin O(100)~nm superconductor patterned on top of a more massive substrate that acts as the dark matter target. Scatters in the substrate produce meV-scale athermal phonons which can break Cooper pairs in the superconducting sensor. 

Energy resolutions of such pairbreaking detectors are often quoted in terms of the ``in-sensor'' resolution on the energy absorbed by a superconducting sensor and the ``in-chip'' resolution on the energy absorbed by the substrate, only a fraction of which makes it into any given superconducting sensor. In-sensor energy resolutions at the 40~meV scale have been achieved in TES-based devices, with best-achieved in-chip energy resolutions at the O(eV)scale~\cite{Fink_2020,WatkinsThesis,RenQET,LargeAreaPhotonDetectorQET}. MKID in-sensor energy resolutions have been achieved at the 2~eV scale~\cite{temples2024performance}, with in-chip energy resolutions reaching down to O(20~eV)~\cite{BULLKID2022,Cardani2018}.

Another candidate for low-threshold sensing uses a superconducting transmon qubit architecture. Commonly used as the fundamental hardware units in a superconducting quantum computer, transmon qubits are anharmonic LC circuits etched into a thin, O(100)~nm superconducting film lying on top of a thicker, O(400)~\(\upmu\)m substrate as shown in Figure \ref{fig:QubitChipSideViewDiagram}. They are composed of a superconducting ``island'' separated from a ground plane by a gap (the capacitor) spanned by a Josephson junction (JJ) or SQUID (the inductor). The flexibility in this device architecture has enabled their use in a variety of quantum sensing applications including single GHz-photon detection~\cite{Dixit}, THz photon detection~\cite{ShawQCD,QCDForTHz}, ultra-sensitive magnetometry~\cite{Magnetometry}, and detection of in-chip charge and phonon bursts~\cite{Wilen}. Many of these sensing modes rely on a demonstrated sensitivity of superconducting qubits to QPs in the leads of the JJ/SQUID\cite{Wang,Downconversion,McEwen,MITRelaxation}, suggesting that qubits may also act as pairbreaking sensors via QP sensing schemes different from (and potentially competitive with) those in other pairbreaking devices. However, only a few studies have been performed to assess the sensitivity of such sensing schemes to in-substrate energy depositions~\cite{SQUAT,KarthikCPAD}.


The objective of this work is to demonstrate a bottom-up, simulation-based estimate of the energy threshold of a phonon-mediated detector built from transmon qubits operated in an ``energy relaxation'' readout scheme. We begin in Section \ref{sec:PhononPropagation} by demonstrating the use of the G4CMP simulation tool in studying phonon transport between an interaction site and a qubit island. Section \ref{sec:SingleQubitResponse} then presents the Quantum Device Response (QDR) simulation tool, which we use to estimate the energy threshold for a single qubit island when using a gate-based ``energy relaxation'' readout scheme~\cite{McEwen,MITRelaxation}. We then explore in Section \ref{sec:ChipResponse} how both chip design and single-qubit response contribute to an overall chip energy threshold. We demonstrate model viability in Section \ref{sec:DataModelComparison} with a preliminary application of our modeling to data recently published in Ref~\cite{MITRelaxation}. We discuss major takeaways from this work in Section \ref{sec:Discussion}, and close in Section \ref{sec:Conclusions}.


\section{Phonon Propagation from an Impact Site to a Qubit Sensor}
\label{sec:PhononPropagation}

\begin{figure}[t!]
\centering
\includegraphics[width=\linewidth]{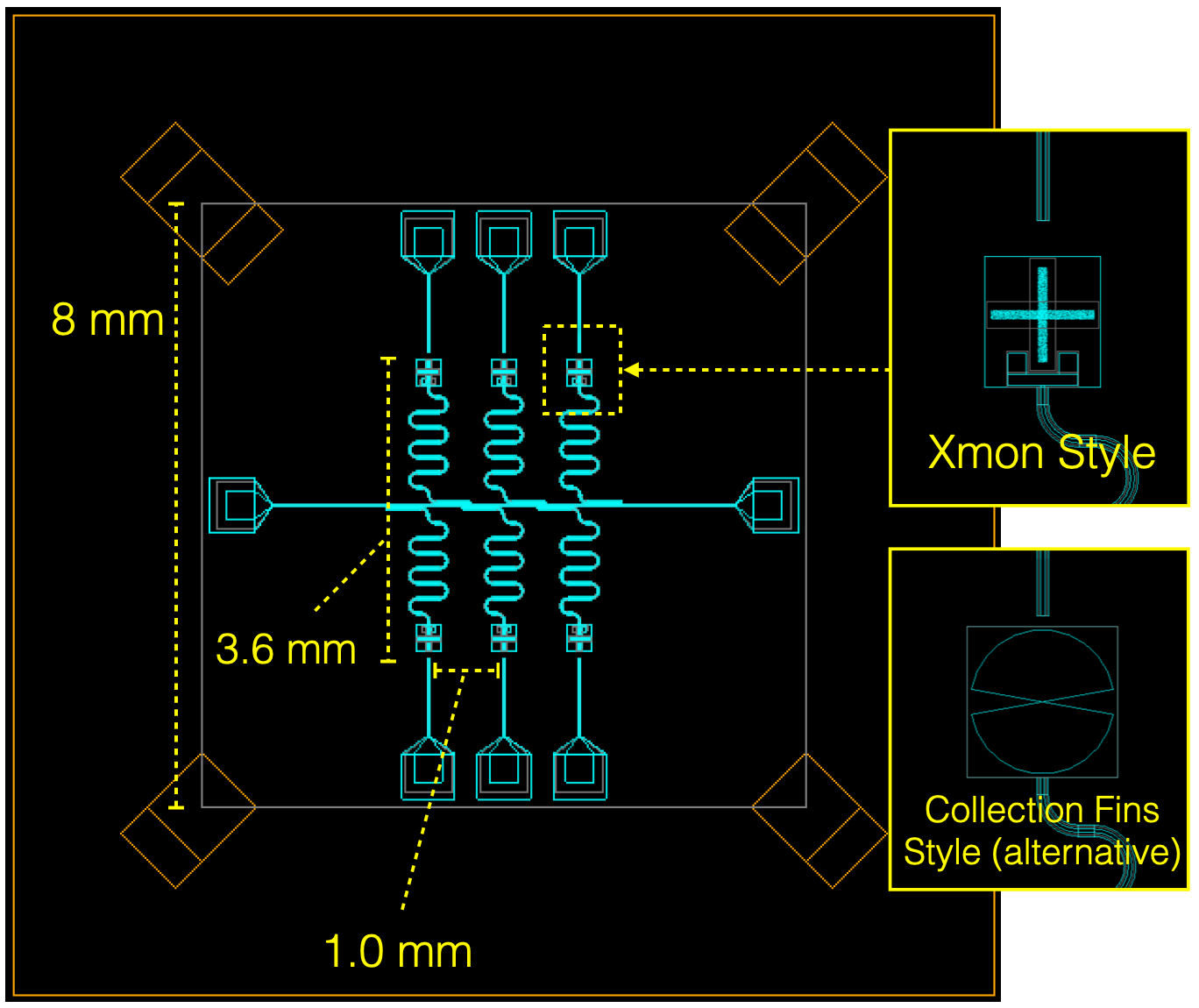}
\caption{Illustration of the baseline 6-qubit chip geometry used in this work's simulation. Each Xmon qubit (top inset) is a small cross (``island'') about 200\(\upmu\)m across, and is coupled to a resonator that is itself coupled inductively to the central feedline. A biasing line and wirebonding pad are included for each qubit. The bottom inset is a visualization of an alternative style of qubit island that employs collection fins, as put forth by Ref.~\cite{SQUAT}. The chip (gray outline) is in contact with copper mounts (orange) on the underside of its four corners.}
\label{fig:QubitChipSimulationRendering}
\end{figure}

The energy threshold of a qubit-based device is strongly dependent on how efficiently energy can propagate from an interaction site to the qubit islands. In this study, we use a Geant4-based particle tracking software called G4CMP~\cite{G4CMP2023} to model this phonon propagation through an example chip geometry (Figure \ref{fig:QubitChipSimulationRendering}): a silicon substrate of dimension 8~mm~x~8~mm~x~380~\(\upmu\)m mounted on four copper corners for thermalization and patterned with six aluminum 100-nm-thick 2D flux-tunable qubits. Each qubit is modeled after the Xmon architecture commonly used for quantum computation, where the cross acts as the qubit island (and in our case, as the target for phonons)~\cite{Xmon}. While this geometry is based on designs from the quantum computing community such as those in Ref.~\cite{Wilen,Downconversion} and is therefore not optimized for particle detection, we use it as a launchpad for discussing more optimal designs for dark matter detection. What follows in this section is functionally an inverse study to Ref.~\cite{Yelton} in which we explore chip parameters to maximize the deposition of phonon energy into the qubit islands.

At an interaction site, energy may be injected into the detector medium in a number of ways: ionization by a high energy particle impact followed by electron-hole phonon scattering and recombination into phonons, direct phonon excitation through a particle impact, or phonon production from stress-induced crystal relaxation~\cite{Wilen,TargetComparison,StressPhononBursts}. For most kinds of interactions that we consider, the majority of the initial energy will be deposited in the device substrate rather than in the much less massive superconducting layer into which the qubits are patterned. Each event type ultimately results in the production of athermal phonons whose total energy is approximately equal to the energy deposited by the initial interaction in the chip.

While the energies of the created athermal phonons will vary depending on the event type, relaxation processes rapidly downconvert (over ns timescales) high-energy phonons into a larger number of lower-energy phonons that travel ballistically in the substrate~\cite{AthermalPhononDownconversionTheory}. In silicon, the rate of this anharmonic downconversion \(\Gamma_{\mathrm{anh}}\) rises with phonon energy to the fifth power, \(\Gamma_{\mathrm{anh}} \propto E_{ph}^{5}\). As a result, the mean free path of this process is approximately 5~\(\upmu\)m for a 30~meV phonon and a much longer 1~cm for a 6~meV phonon. Similarly, the rate at which phonons scatter on isotopic impurities within the crystal lattice scales as \(E_{ph}^{4}\), increasing steeply with phonon energy. When the mean free paths for both of these processes are longer than the device dimensions, the phonons will stream largely unimpeded along the material's phonon caustics~\cite{EveryCaustics}, and their momenta and energy spectra are influenced primarily by boundary scatters. For our baseline detector which has a 380~\(\upmu\)m thickness (or for detectors with similar thickness), phonons are effectively ballistic at energies below the 6-meV scale. These low-energy phonons are responsible for conveying the deposited energy from the interaction site to the qubit sensors, and do so quasidiffusively over 10-100~\(\upmu\)s timescales for standard 1x1-cm\(^{2}\)-area wafers~\cite{Wilen}. 

Phonons may be lost either through absorption in the superconducting layer or by absorption in the copper thermalization mounts in the corners on the underside of the chip. In this work, we model an ``all-or-none'' phonon absorption process at a given surface with an effective ``phonon absorption probability'' (\(p_{a}\)), such that a phonon incident upon an interface has a probability \(p_{a}\) of being fully absorbed and a  probability \(1-p_{a}\) of being fully reflected. We will use the notation \(p_{a,s}\) to represent this probability for interfaces between substrate and superconductor, and \(p_{a,c}\) to represent this probability for interfaces between substrate and (copper) corner thermalization mounts. This ``all-or-none'' model is a necessary oversimplification of the physics involved in phonon absorption, as rigorous first-principles modeling needs to capture complex dependencies on several parameters such as phonon energy and angle, interfacial acoustic mismatch, film crystallinity and thickness, and (in the case of superconductors) the gap parameter \(\Delta\), not all of which are currently captured in G4CMP~\cite{KaplanAcousticMatching,McEwen}. Some of this complexity can be mapped into our simple modeling paradigm using known approximations to \(p_{a,s}\):
\begin{equation}
\label{eq:paVsEph}
p_{a,s} \simeq 1 - \exp\Bigg[-\frac{2l}{\pi v_{s} \tau^{ph}_{0}}\Big(\frac{E_{ph}}{\Delta}\Big)\Bigg],
\end{equation}
where \(E_{ph}\) is the phonon energy, \(l\) is the film thickness, \(\tau_{0}^{ph}\) is a material-dependent characteristic phonon scattering time~\cite{Kaplan}, \(v_{s}\) is the sound speed in the material, and \(\Delta\) is the superconducting gap energy. This expression holds in the limit \(E_{ph}\gg\Delta\) \cite{McEwen,Yelton}, which is a reasonable assumption for a phonon created in a downconversion process in a silicon chip and impinging upon a superconducting aluminum film. For a film thickness of 100~nm, Equation~\ref{eq:paVsEph} gives \(p_{a,s}\gtrsim0.2\) for \(E_{ph}>1\)~meV. We will therefore focus our study on \(p_{a,s}\) values in the range from 0.1 to 1.0. A more rigorous handling of superconductor and normal-metal phonon modeling will be the focus of follow-up work.


\begin{table*}[t!]
\centering
\caption{Estimates of \(\eta_{ph,sp}\) and \(E_{thr,chip}\) for a set of 12 chip designs. Relative statistical uncertainties on the simulated estimates of \(\eta_{ph,sp}\) are at or below the few-\(\%\) level. The calculation of thresholds in the rightmost column relies on the discussion in Sections~\ref{sec:SingleQubitResponse} and \ref{sec:ChipResponse}, but results are tabulated here for conciseness. For threshold estimates, the single-qubit readout parameters used are the near-term ``baseline'' ones discussed: \(T_{1,base}=\)2~ms, \(\mathcal{F}=0.98\) (see Section~\ref{sec:SingleQubitResponse} text for definitions). Under these conditions, \(\sigma_{E,\mathrm{abs}}\simeq0.088\)eV. We note that for Designs 13 and 14, the calculation of threshold accounts for a few additional effects that modify \(\sigma_{E,\mathrm{abs}}\) as discussed at the end of Section~\ref{subsec:SensorEnergyResolutionAndThreshold}.}
\begin{tabular}{|p{1.1cm}||p{1.4cm}|p{2.2cm}|p{1.2cm}|p{2.4cm}|p{2.4cm}||p{3.1cm}|p{1.6cm}|}
\hline
Design & Number of Qubits & Qubit Design & Ground Plane & \text{Si-SC phonon} absorption prob. & \text{Si-Cu phonon} absorption prob. & Spatially-averaged phonon collection eff. & Chip Threshold \\
 & \(N_{q}\) & & & \(p_{a,s}\) & \(p_{a,c}\) & \(\eta_{ph,sp}\) & \(E_{thr,chip}\) \\
\hline
\hline
1 & 6 & Xmon & Full & 1.0 & 0.1 & 0.14\(\%\) & 737~eV \\
2 & 6 & Xmon & Full & 0.1 & 0.1 & 0.12\(\%\) & 860~eV \\
3 & 6 & Xmon & Limited & 1.0 & 0.1 & 2.07\(\%\) & 49~eV \\
4 & 6 & Xmon & Limited & 0.1 & 0.1 & 1.44\(\%\) & 71~eV \\
\hline
5 & 6 & Xmon & Full & 1.0 & 1.0 & 0.14\(\%\) & 737~eV \\
6 & 6 & Xmon & Full & 0.1 & 1.0 & 0.12\(\%\) & 860~eV \\
7 & 6 & Xmon & Limited & 1.0 & 1.0 & 1.76\(\%\) & 58~eV \\
8 & 6 & Xmon & Limited & 0.1 & 1.0 & 0.76\(\%\) & 135~eV \\
\hline
9 & 2 & Xmon & Full & 1.0 & 0.1 & 0.05\(\%\) & 1157~eV \\
10 & 10 & Xmon & Full & 1.0 & 0.1 & 0.24\(\%\) & 574~eV \\
11 & 2 & Xmon & Limited & 1.0 & 0.1 & 1.38\(\%\) & 41~eV \\
12 & 10 & Xmon & Limited & 1.0 & 0.1 & 2.39\(\%\) & 57~eV \\
\hline
13 & 6 & Collection Fins & Limited & 1.0 & 0.1 & 17.0\(\%\) &  O(0.1)~eV \\
14 & 6 & Collection Fins & Limited & 0.1 & 0.1 & 12.6\(\%\) &  O(0.1)~eV \\
\hline
\end{tabular}
\label{table:ConfigurationsVsThresholds}
\end{table*}
\vspace{5mm}

\begin{figure}[t!]
\centering
\includegraphics[width=\linewidth]{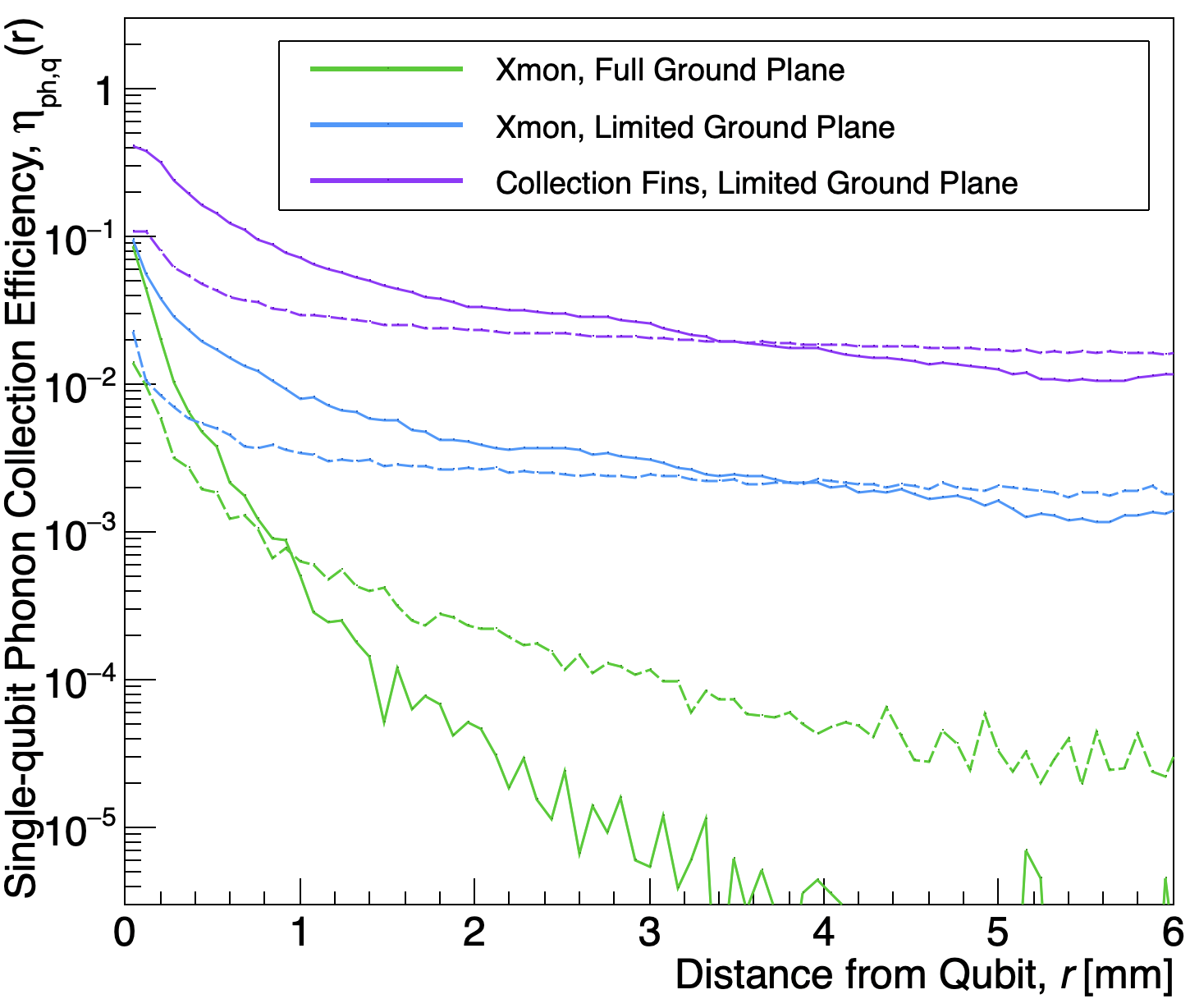}
\caption{Single-qubit phonon collection efficiency \(\eta_{ph,q}\) versus radial distance from a given qubit island. Each curve is averaged over the six qubits simulated in our baseline geometry. This simulation is estimated for six different chip designs: full ground plane, limited ground plane, and limited ground plane with collection-fin qubit design. Solid (dashed) lines are for a \(p_{a,s}\) of 1.0 (0.1) at the Si-SC interface.}
\label{fig:PhononAbsorptionEfficiency}
\end{figure} 


With G4CMP, we first estimate the probability with which a ballistic phonon emerging from an interaction point will be absorbed into a qubit island. We define a qubit's spatially-varying phonon collection efficiency \(\eta_{ph,q}(r)\) as the average fraction of a total emitted phonon energy that is absorbed by a qubit a distance \textit{r} from the emission point. We launch simulated 20-meV phonons uniformly throughout the chip, explicitly tabulate this fraction as a function of radial distance from each qubit, and average the fraction across the six qubits (Figure~\ref{fig:PhononAbsorptionEfficiency}). We run these simulations for three different potential chip conditions that vary in qubit design and overall superconductor coverage: an Xmon-style island with a full Al ground plane, an Xmon-style island with a nearly absent (``Limited'') ground plane, and a collection-fin style island (Figure~\ref{fig:QubitChipSimulationRendering} lower inset) with a nearly absent ground plane.\footnote{While we present this third design to highlight its advantageous phonon collection, we have not yet performed studies to understand how feasibly this design may function as a transmon from an RF standpoint. For more discussion, see Ref.~\cite{SQUAT}} In the three above-mentioned cases, the ratios of total qubit area to total superconductor area are 0.0011, 0.019, and 0.156, respectively. For each of these, we make estimates for \(p_{a,s}=0.1\) and \(p_{a,s}=1.0\) at the Si-SC (silicon-superconductor) interface, and assume \(p_{a,c}=0.1\) at the Si-Cu interface where the chip's four corners sit on the copper thermalization mount.

These six scenarios in Figure~\ref{fig:PhononAbsorptionEfficiency} already provide a wide range of possible phonon collection efficiencies. A chip with a full ground plane may display phonon collection efficiencies that are reduced by orders of magnitude relative to a chip with a spatially limited ground plane. A full ground plane significantly decreases the ratio of ``sensing'' superconductor (i.e. the qubit island) area to total superconductor area in which phonons may be absorbed, reducing the overall collection efficiency. An additional order of magnitude improvement in phonon collection can be achieved with the use of collecting fins for the same reason. For a given ground plane scenario, a lower probability of phonon absorption at the Si-SC interface will lead to a lower \(\eta_{ph,q}\) close to the qubit, but a larger \(\eta_{ph,q}\) far from the qubit. This is as a result of the increased ability of phonons to travel laterally in the plane of the chip by reflecting multiple times off of its top and bottom.

Another useful metric to quote is the total phonon collection efficiency spatially averaged over the entire chip, \(\eta_{ph,sp}\), which we define as
\begin{equation}
\eta_{ph,sp} \equiv \biggl< \frac{\sum_{i}E_{\mathrm{dep},i}}{E_{\mathrm{dep,chip}}}\biggr>,
\end{equation}
where \(E_{\mathrm{dep},i}\) is the energy absorbed by qubit \(i\), \(E_{\mathrm{dep,chip}}\) is the total deposited energy, and the brackets denote spatial averaging of the location of the energy deposition \(E_{\mathrm{dep,chip}}\) over the chip. This is a single metric that describes the overall conversion of in-chip energy to in-qubit energy. Table \ref{table:ConfigurationsVsThresholds} presents estimates of this spatially averaged \(\eta_{ph,sp}\) for an expanded set of simulation conditions. Beyond the chip conditions shown in  Figure~\ref{fig:PhononAbsorptionEfficiency} (Designs 1-4,13,14), we explore wide variations in the absorption probability for phonons impinging upon the Si-Cu interface at the mount points in the four corners of the chip (Designs 5-8), as this probability is not easy to know \textit{a priori}. We also study variations in the number of qubit-resonator-control-line assemblies included (Designs 9-12), to understand how the overall collection scales as phonon energy is necessarily split between multiple sensors.

An analytical approximation to \(\eta_{ph,sp}\) can be constructed as the ratio of sensitive to total absorptive area weighted by the phonon absorption probability at each interface~\cite{SQUAT}:
\begin{equation}
\label{eq:EtaAnalyticalGeneral}
\eta_{ph,sp} \approx \frac{p_{a,s}A_{s}}{\sum_{i}p_{a,i}A_{i}},
\end{equation}
where \(A_{s}\) is the sum of areas of all sensitive elements (qubit islands) and \(A_{i}\) represents a more general interfacial element's area, whether sensitive or not. We recast this general form slightly to acquire a form explicitly dependent on the number of qubits \(N_{q}\):
\begin{equation}
\label{eq:EtaAnalyticalSpecific}
\eta_{ph,sp} \approx \frac{N_{q}A_{q}p_{a,s}}{N_{q}A_{t}p_{a,s}+A_{c}p_{a,c}+f(N_{q})},
\end{equation}
where the \(A_{q}\) is area of a qubit island, \(A_{t}\) is the total superconducting interfacial area added per qubit assembly (i.e. qubit, resonators, control lines, etc.), \(A_{c}\) is the total area of the copper corner thermal mounts, and \(f(N_{q})\) is a function representing the average product of absorption probability and area for the remaining superconductor area. The form of \(f\) is dependent on ground plane scenario: for a negligible ground plane, \(f(N_{q})\) is just the product of the transmission line area and \(p_{a,s}\), and is therefore independent of \(N_{q}\). For a full ground plane, \(f(N_{q}) \simeq p_{a,s}(A_{chip} - N_{q}A_{t})\), and therefore falls with increasing \(N_{q}\). We find that Equation~\ref{eq:EtaAnalyticalSpecific} agrees with our simulation-determined \(\eta_{ph,sp}\) values to within approximately 15\(\%\) across the design variations in Table~\ref{table:ConfigurationsVsThresholds}.


We can intuit a few key lessons from the combination of Equation \ref{eq:EtaAnalyticalSpecific} and Table \ref{table:ConfigurationsVsThresholds}. First, the clearest advantage in overall phonon collection unsurprisingly comes from a limited ground plane, a large \(p_{a,s}\), and from use of a qubit architecture with collection fins. In this scenario, the overall energy collection rises with the number of qubits but the energy collected per qubit in any given event falls, as the total phonon energy must be shared among a larger number of absorbers. Second, the effect of varying the probability of phonon absorption at the thermal mount interface is strongest in the limited ground plane case with a Si-SC absorption probability of 0.1: if phonon loss is weaker in the superconductor, thermal mount losses have a stronger effect on the overall phonon loss.

\section{Single Transmon Signal Response to Substrate Phonons}
\label{sec:SingleQubitResponse}

The second component responsible for determining an overall detector energy threshold is the signal response of an individual qubit island to incident phonon energy. This depends on the time evolution of the in-qubit QP density, the evolution of the resulting quantum state, and the scheme used to read out that quantum state. The estimate of threshold also depends on how interactions are reconstructed using this readout scheme. In this section we discuss in detail both how we model the system time evolution and how we reconstruct an energy deposition in a qubit. We then present estimates of our single-qubit threshold for a variety of qubit performance parameters.

As the parameter space affecting signal response is very broad, we include Table~\ref{table:ConstantQDRSimulationParameters} as a guide to the single-qubit response parameters that we keep constant throughout this process for the purpose of reducing the dimensionality of our simulation parameter space. Notably, these parameters are for the Xmon-style qubit design, as the collection fin designs (Designs 13 and 14 in Table \ref{table:ConfigurationsVsThresholds}) have yet to be experimentally demonstrated.
Further discussion of these parameters can be found in the sections that follow.

\subsection{Modeling Qubit Signal Time Evolution}
\label{subsec:QubitTimeSignalEvolution}

As G4CMP does not yet handle the QP/superconductor response rigorously in arbitrary superconducting layers, a different tool is needed to properly capture this physics in this modeling step. To do this, we have developed a simulation tool called QuantumDeviceResponse (QDR). QDR takes as input the timestamps and magnitudes of energies absorbed in a qubit island simulated in G4CMP, and performs three operations on them:
\begin{enumerate}
\item It converts the deposited energy into a QP density, and evolves this QP density in discret time steps \(\delta t\) according to established models of QP dynamics in superconductors.
\item Using a simplified representation of qubit state, it uses the QP density at a given time to stochastically determine qubit state evolution.
\item It simulates qubit control and measurement using simplified gate sequences to mimic a gate-based readout scheme. 
\end{enumerate}
In the following discussions we explore these operations in more detail.

\subsubsection{Quasiparticle Creation and Evolution}
\label{subsubsec:QuasiparticleCreationAndEvolution}

\begin{table}[t!]
\centering
\caption{QDR parameters held constant within the simulation study.}
\begin{tabular}{|p{1.45cm}||p{3.6cm}|p{1.7cm}|p{1.2cm}|}
\hline
Parameter & Description & Value & Refs. \\
\hline
\hline
Material & SC material & Aluminum &  \\
\(\delta t\) & QDR time step & 100~ns & \\
\(r\) & Recombination rate & 0.005 ns\(^{-1}\) & \cite{Wang,Kaplan} \\
\(s_{0}\) & Linear loss rate & 10\(^{-6}\) ns\(^{-1}\) & \cite{Wang} \\
\(g\) & QP generation rate & 10\(^{-14}\) ns\(^{-1}\) & \cite{Wang} \\
\(n_{CP}\) & Superfluid (CP) density & 4\(\times 10^{24}\)~m\(^{-3}\) & \cite{Wang,McEwen} \\
\(\Delta\) & SC gap energy & 180~\(\upmu\)eV & \cite{Wang,McEwen} \\
\(V\) & Island volume & 1000~\(\upmu m^{3}\) &  \\
\(\epsilon\) & Phonon-to-QP efficiency & 0.6 & \cite{MartinisSaving} \\
\(F\) & QP-creation Fano factor & 0.2 & \cite{FanoFactor} \\
\(\omega_{q}\)/2\(\pi\) & Qubit frequency & 4~GHz & \\
\(N\) & Measurement binning & 20 & \\

\hline
\end{tabular}
\label{table:ConstantQDRSimulationParameters}
\end{table}
\vspace{5mm}

Once in the qubit island of volume \(V\), phonons with energy larger than twice the superconducting gap, 2\(\Delta\), can break Cooper pairs (CPs), producing Bogoliubov QPs. QPs with energies higher than \(\Delta\) lose energy by radiating additional phonons in the superconductor, which if sufficiently energetic can create additional QP pairs. At the end of such a cascade, the final state of the detector involves a set of created QP pairs each with near-\(\Delta\) energies and a set of low-energy phonons that may re-enter the substrate. The fraction \(\epsilon\) of an initial energy deposition retained in QPs is dependent on the initial phonon energy and film. For Al, where 2\(\Delta\sim360~\upmu\)eV is much smaller than the few-meV energy of an average ballistic phonon, \(\epsilon\simeq 0.6\)~\cite{MartinisSaving}. In QDR, an energy deposition \(E_{\mathrm{dep}}\) produces a number of QPs \(N_{qp}\) according to a gaussian distribution with a mean of \(E_{\mathrm{dep}}\epsilon/\Delta\) and a spread \(\sqrt{FE_{\mathrm{dep}}\epsilon/\Delta}\), where a Fano factor \(F=0.2\) has been used~\cite{FanoFactor}. For ballistic phonons arriving later at the qubit over approximately 10-100~\(\upmu\)s, this overall \(N_{qp}\) will increase in discrete steps over this timescale.

\begin{figure}[t!]
\centering
\includegraphics[width=\linewidth]{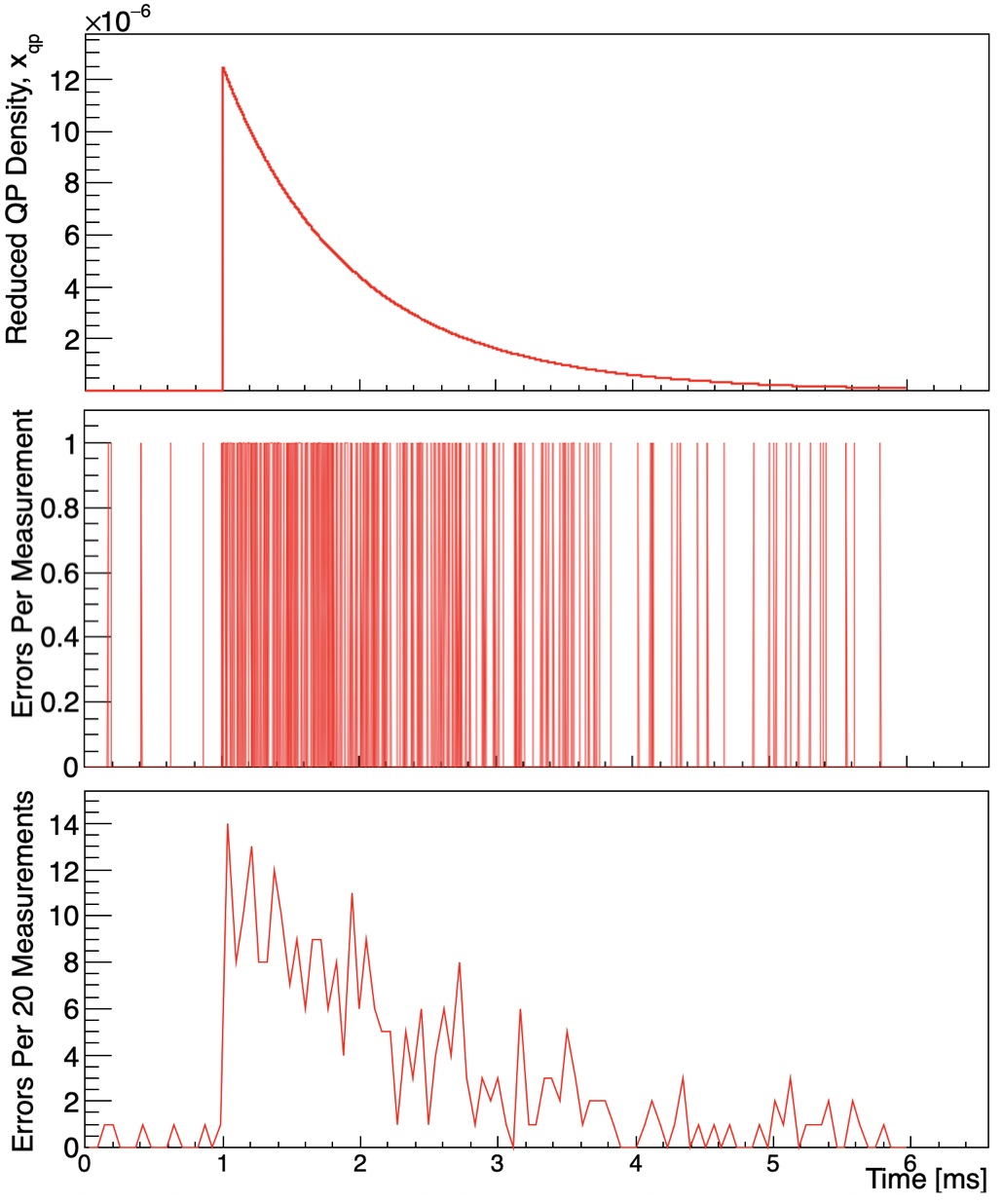}
\caption{A simulated event in which 15~eV is deposited directly into a single transmon island at \(t=\)1~ms. \textbf{Top:} Reduced QP density as a function of time. \textbf{Center:} The time series of read-out qubit state for the ``energy relaxation'' readout scheme in Figure \ref{fig:McEwenReadoutScheme}. Here, measurements right after energy injection frequently return an ``error'', and measurements many~ms after the injection less frequently return an ``error'' due to the reduced QP density. For this readout scheme, we assume a qubit \(T_{1,\mathrm{base}}=~\)2~ms, \(\mathcal{F}=98\%\), and search time \(\Delta t=2000\)~ns. \textbf{Bottom:} The errors in the central curve, binned in time such that each ``sample'' represents the number of errors in 20 measurements of the qubit.}
\label{fig:ExampleWaveform}
\end{figure}

After an energy deposition in the qubit, evolution of QP populations is governed primarily by diffusion, recombination, and trapping. Diffusion throughout islands of order 200~\(\upmu\)m happens within O(10)~\(\upmu\)s~\cite{DiffusionEnergyDependence,DongDiffusion,QPDiffusionHsieh}, allowing us to treat QPs effectively as if they exist with a uniform density \(n_{qp} \equiv N_{qp}/V\) in the island and leads after this time has elapsed. Knowing the superfluid (CP) density \(n_{CP}\) for a given material, it is then possible to model the normalized quasiparticle density \(x_{qp} \equiv n_{qp}/n_{CP}\) with
\begin{equation}
\label{eq:QPDiffEq}
\frac{dx_{qp}}{dt} = -rx_{qp}^{2} - s_{0}x_{qp} + g.
\end{equation}
where the coefficients \(r\) and \(s_{0}\), and the constant \(g\) represent rates of QP ``self''-recombination, linear loss due to QP trapping or recombination with ``quiescent'' QPs, and QP generation, respectively~\cite{Wang,ExcessQuasiparticlesMKID,temples2024performance}. In each time step \(\delta t\), QDR computes the corresponding \(dx_{qp}\) using these parameters and uses it to update the total \(x_{qp}\) in a simulated qubit (Figure \ref{fig:ExampleWaveform}, top).


For our model, we motivate choices of \(r\), \(s_{0}\), and \(g\) using values measured in literature. The recombination coefficient \(r\) is expected to be as large as its theoretically motivated value, \(r_{\mathrm{Kaplan}}=0.05\)~ns\(^{-1}\), but may be suppressed by around an order of magnitude or more depending on phonons that re-break CPs before leaving the film~\cite{Wang,Kaplan,Yelton}. While this suppression factor is device-specific, we have also observed that the results that follow are only weakly dependent on \(r\) within the range of values typically measured. In this study we use a value of $r=$~0.005~ns\(^{-1}\), corresponding to a suppression factor of 10. The values of \(s_{0}\) and \(g\) are difficult to predict \textit{a priori} and depend heavily on experimental parameters like material impurities, spatial variation in the superconducting gap \(\Delta\), and IR shielding~\cite{qTLS,PAPS}. In our modeling, these are set to the values in Table \ref{table:ConstantQDRSimulationParameters}.

\subsubsection{Quantum State Evolution}
\label{subsubsec:QuantumStateEvolution}

QDR contains a description of a qubit state that also evolves in time. For this study, we use a very simple model in which the state is in either the first excited energy eigenstate \(\ket{1}\) or the energy ground state \(\ket{0}\). The state can be changed using simulated gates in a gate sequence: a \(\pi\)-pulse performs a bit flip operation, taking \(\ket{0}\rightarrow\ket{1}\) and \(\ket{1}\rightarrow\ket{0}\). The state can also be changed if energy relaxation occurs. QPs present in the qubit island may tunnel across the qubit's Josepshon junction and induce relaxation of a qubit from the excited state to ground state. The rate of relaxations due to such tunneling events in a qubit with frequency \(\omega_{q}\) is known~\cite{CatelaniFrequencyShifts} to be
\begin{equation}
\label{eq:TimeIndependentDecoherence}
\Gamma_{qp} = \sqrt{\frac{2\omega_{q}\Delta}{\pi^{2}\hbar}}x_{qp},
\end{equation}
which can be recast in terms of a characteristic relaxation coherence time \(T_{1,qp}\) = \(1/\Gamma_{qp}\). The relaxation rate contributes to an overall energy relaxation rate of a qubit \(\Gamma_{\mathrm{tot}} = \Gamma_{\mathrm{other}} + \Gamma_{qp}\), where \(\Gamma_{\mathrm{other}}\) represents the energy decay rate contributions from non-QP-related dissipative effects caused by other environmental noise~\cite{WangMultiChip}.

QDR computes a total relaxation rate from both the instantaneous QP density at a given time and a parameterized ``baseline'' coherence time, \(T_{1,\mathrm{base}}\), due to non-QP effects. If the qubit is brought away from the \(\ket{0}\) state, QDR uses the instantaneous characteristic decay time \(T_{1}\) to randomly sample the next time at which state relaxation back to \(\ket{0}\) is induced. It updates this estimate on the fly as \(T_{1,qp}\) evolves, using techniques similar to those used in Ref.~\cite{SLACGeant4} for particle tracking through inhomogeneous materials.

\subsubsection{Quantum State Readout}
\label{subsubsec:QuantumStateReadout}

The final role of QDR is to simulate the readout value extracted from a qubit. We include a projective measurement as a ``readout gate'' that can map the qubit state \(\ket{0}\) (\(\ket{1}\)) onto a boolean 0 (1) value. This gate suffers from readout noise, characterized by a probability \(p_{01}\) of incorrectly measuring a true \(\ket{1}\) state as \(\ket{0}\) and a probability \(p_{10}\) of incorrectly measuring a true \(\ket{0}\) state as \(\ket{1}\). The single-shot-fidelity (\(\mathcal{F}\)) of this readout gate is then defined~\cite{SSFDefinition} as
\begin{equation}
\label{eq:SSFDefinition}
\mathcal{F} = 1 - p_{10} - p_{01}.
\end{equation}
For this work, we make the simplifying assumption that \(\mathcal{F}\) is symmetric, i.e. \(p_{10} = p_{01}\).

While for now the quantum state evolution and readout simulated in QDR is limited to the operations and physics listed above, a thrust of future effort is its expansion to handle the modeling of the charge parity state of a qubit, \(\pi/2\)-pulses, and phase noise in addition to relaxation noise.\footnote{More generally, QDR is designed to facilitate construction of the readout response for general pairbreaking sensors and modeling of chips with multiple sensor types integrated. To that effect, it also includes a response for phonon-sensing microwave kinetic inductance detectors (MKIDs).}

\begin{figure}[t!]
\centering
\includegraphics[width=\linewidth]{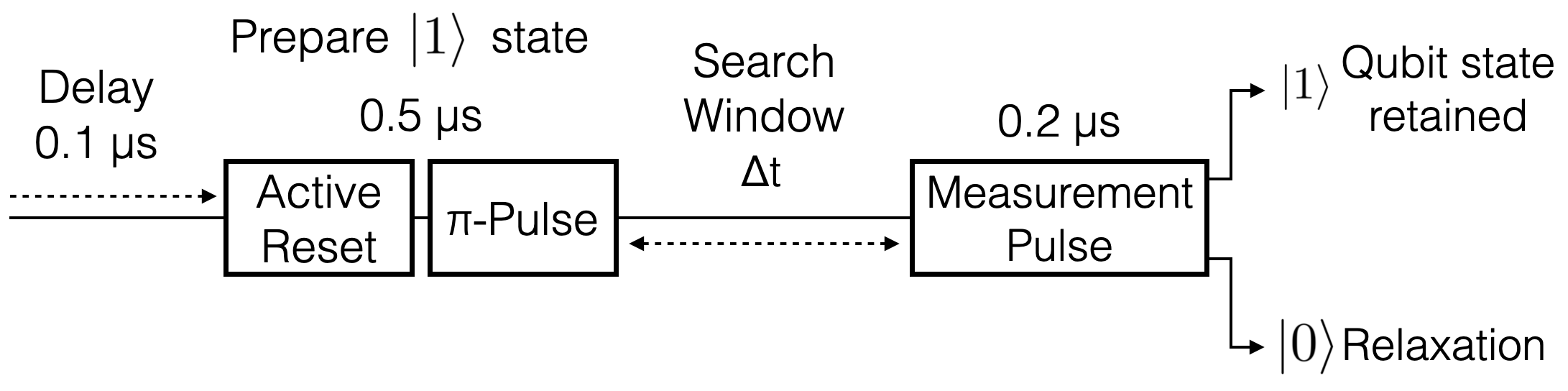}
\caption{The qubit gate sequence representing one measurement forming the foundation of the energy relaxation detection scheme. \(N\) such measurements are binned into one ``waveform sample.''}
\label{fig:McEwenReadoutScheme}
\end{figure}

\subsection{Reconstruction of an In-Qubit Energy Deposition Using an Energy Relaxation Readout Scheme}
\label{subsec:ReconstructionOfAnInQubitEnergyDeposition}

Determining the energy threshold achievable by a qubit sensor requires defining a readout scheme and signal reconstruction methodology. Here, we focus on a time-binned ``energy relaxation'' scheme (Figure~\ref{fig:McEwenReadoutScheme}) used in recent studies of superconducting qubit susceptibility to environmental radiation~\cite{McEwen,MITRelaxation}, using QDR to directly simulate indefinitely-repeated calls to the gate sequence. This scheme is functionally equivalent to measuring a change in coherence time \(T_{1}\) as a function of the time since an energy deposition. We first build a signal template for waveforms read out with this scheme, and then discuss a strategy for energy reconstruction using that signal template.

\subsubsection{Signal Template for Energy Relaxation Readout Scheme}
\label{subsubsec:SignalTemplate}

We now construct a template \(S(t)\) for the signal read out using this relaxation scheme after an energy deposition in a qubit island. The output of one cycle of this gate sequence is a boolean value (Figure~\ref{fig:ExampleWaveform}, center) which reflects whether a decoherence has happened within the search window \(\Delta t\) occurring after the \(\pi\)-pulse. These boolean values can be binned in time to produce a ``waveform'' of qubit errors per \(N\) measurements (Figure~\ref{fig:ExampleWaveform}, bottom). Unless otherwise noted, we take \(\Delta t=2~\upmu\)s and \(N=20\) so that we may still maintain sufficient time granularity to make use of the shape of a waveform in energy reconstruction.

First, we assume an initial instantaneous energy deposition \(E_{\mathrm{dep}}\) causes an initial QP density \(x_{i}\) in an island volume \(V\). For this scenario, the solution to the Equation~\ref{eq:QPDiffEq}, given in Ref.~\cite{Wang}, is
\begin{equation}
\label{eq:xqpSolution}
\begin{split}
x_{qp} = & x_{i}\frac{1-r'}{e^{t/\tau_{ss}}-r'}+x_{0} \\
= & \Big[\frac{E_{\mathrm{dep}}\epsilon}{n_{CP}V\Delta}\Big]\frac{1-r'}{e^{t/\tau_{ss}}-r'}+x_{0}. \\
\end{split}
\end{equation}
Here, \(\tau_{ss}\), \(r'\), and \(x_{0}\) are functions of \(r\), \(s_{0}\), \(g\), and \(x_{i}\).\footnote{Though not explored in this work, we note that phonon ``recycling'' from \(\simeq\)2\(\Delta\) recombination phonons may distort this functional form: 2\(\Delta\) phonons can rebreak CPs, which may later recombine to produce more 2\(\Delta\) phonons. This cycle may continue as long as the 2\(\Delta\) phonons do not exit through thermal contacts. As a result, for chips with minimal contact with their thermal mounts, this may significantly lengthen the time dependence of \(x_{qp}\) and the overall response time of the qubits~\cite{McEwen}.} Notably, \(\tau_{ss}\) is the ``steady-state'' decay time for the QP relaxation, found when \(x_{qp}\) becomes small enough that self-recombination becomes subdominant.

The rate of qubit decoherence then follows from Equation~\ref{eq:TimeIndependentDecoherence} with an additional non-QP-induced term:
\begin{equation}
\label{eq:timeDependentDecoherence}
\Gamma(t) = \sqrt{\frac{2\omega_{q}\Delta}{\pi^{2}\hbar}}\Bigg(\Big[\frac{E_{\mathrm{dep}}\epsilon}{n_{CP}V\Delta}\Big]\frac{1-r'}{e^{t/\tau_{ss}}-r'}+x_{0}\Bigg) + \Gamma_{\mathrm{other}}.
\end{equation}
Using our readout scheme, this time-varying relaxation rate can be effectively mapped onto a probability of measuring a readout value of 0 (an ``error'') after the search window \(\Delta t\) has elapsed. Using the substitution \(r' = \tau_{ss}x_{i}r / [\tau_{ss}x_{i}r + 1]\) from Ref.~\cite{Wang} (with \(x_{i}\) defined above), and consolidating variables by defining \(\alpha~=~\Delta t\sqrt{2\omega_{q}\Delta/\pi^{2}\hbar}\), \(\beta=n_{CP}V\Delta/\epsilon\), and \(\gamma=-\alpha x_{0} - \Gamma_{\mathrm{other}}\Delta t\), we find that the probability \(p_{r}\) of a qubit relaxing in a time \(\Delta t\) between a \(\pi\)-pulse and measurement pulse on any given measurement is then given by:
\begin{equation}
\label{eq:pr}
\begin{split}
p_{r} =\ & 1-e^{-\Gamma(t)\Delta t} \\
=\ & 1-\exp\Bigg[\frac{-\alpha E_{\mathrm{dep}}}{E_{\mathrm{dep}}[\tau_{ss}r(e^{t/\tau_{ss}}-1)] + \beta e^{t/\tau_{ss}}}-\gamma\Bigg].
\end{split}
\end{equation}
Here, \(\gamma\) represents a ``baseline'' contribution to relaxation composed of both baseline QP contributions (\(\alpha x_{0}\)) and non-QP contributions (\(\Gamma_{\mathrm{other}}\Delta t\)).

The probability \(p_{\mathrm{obs}}\) of acquiring 0 (an error) in a single measurement depends on \(p_{r}\) and any imperfect single-shot fidelity in the qubit and readout chain. Using the nomenclature of Section~\ref{subsubsec:QuantumStateReadout} and defining \(p_{0}\) as the probability of correctly measuring a true \(\ket{0}\) state as \(\ket{0}\), we find
\begin{equation}
\label{eq:pObs}
\begin{split}
p_{\mathrm{obs}} =\ & p_{r}p_{0} + (1-p_{r})p_{01} \\
=\ & p_{r}(p_{0}-p_{01}) + p_{01} \\
=\ & p_{r}(1-p_{10}-p_{01}) + p_{01} \\
=\ & p_{r}*\mathcal{F} + p_{01} \\
\end{split}
\end{equation}
Summing \(N\) sequential measurements in a time ``bin'' gives our signal template \(S(t)\). The value of such a waveform sample follows a binomial distribution described by \(N\) Bernoulli trials and a success probability \(p_{\mathrm{obs}}\). The expected signal template in this readout scheme is then simply
\begin{equation}
\label{eq:FullSignal}
\begin{split}
S(t) =\ & Np_{\mathrm{obs}} \\
=\ & N(p_{r}\mathcal{F} + p_{01}) \\
=\ & N\Bigg(\mathcal{F}\Big(p_{r}-\frac{1}{2}\Big)+\frac{1}{2}\Bigg),
\end{split}
\end{equation}
where in the last expression we have taken advantage of our assumption of a symmetric \(\mathcal{F}\). This expression is valid for values of \(N\) for which the relaxation observation probability \(p_{\mathrm{obs}}\) stays relatively constant over one \(N\)-measurement sample. In the following development, we limit ourselves to using \(N\) and other parameters such that this \(p_{\mathrm{obs}}\) is constant to well within \(10\%\) in such a sample.

\subsubsection{Energy Reconstruction Methodology}

Using the signal template in Equation~\ref{eq:FullSignal}, we may estimate an in-qubit energy deposit from a binned time series of qubit measurements. The following reconstruction procedure is based loosely on the optimal filter strategy employed by Ref.~\cite{temples2024performance,GolwalaThesis}, though due to the non-gaussianity of the signal under consideration here, we must use explicit maximum likelihood fits in our reconstruction methodology to extract physics information from a waveform. We will first briefly describe this general fitting procedure as it pertains to our binomially-distributed signal template \(S(t)\).

For any single waveform sample built from \(N\) measurements each with probability \(p_{\mathrm{obs}}\) of yielding an error, the probability of finding \(s\) errors is given by a binomial distribution:
\begin{equation}
\label{eq:SingleSampleProbability}
p(s) = \binom{N}{s}p_{\mathrm{obs}}^{s}(1-p_{\mathrm{obs}})^{(N-s)}
\end{equation}
Since the expected time-dependent waveform composed of many \(N\)-measurement samples is given by the signal template \(S(t)=Np_{\mathrm{obs}}(t)\) in Equation~\ref{eq:FullSignal}, the joint probability of acquiring a set of \textit{measured} waveform samples \(\{s_{i}\}\) at times \(\{t_{i}\}\) given this expected signal template can be written as:
\begin{equation}
\label{eq:MultiSampleProbability}
p(\{s_{i}\}|S(t)) = \prod_{i}\binom{N}{s_{i}}p_{\mathrm{obs}}^{s_{i}}(t_{i})(1-p_{\mathrm{obs}}(t_{i}))^{(N-s_{i})}\\
\end{equation}
where the range of times \(\{t_{i}\}\) should be between the injection time \(t_{0}\) and a few characteristic decay times after \(t_{0}\). The probability in Equation~\ref{eq:MultiSampleProbability} is also equivalent to the likelihood \(\mathcal{L}\) of the signal template given the measured waveform \(\{s_{i}\}\). Maximizing the likelihood as a function of a signal template parameter of interest (with all others constrained) gives a best-fit value of that parameter of interest. We will use this maximum-likelihood (ML) fitting process twice in our energy reconstruction methodology, as discussed below.

First, a detector calibration must be performed to acquire the pulse-shape-determining quantities \(\tau_{ss}\), \(r\), and \(\gamma\). The quantity \(\gamma\) can be extracted from basic qubit \(T_{1}\) and \(\mathcal{F}\) characterization tests, and so does not require us to use the ML formalism. However, the quantities \(\tau_{ss}\) and \(r\) are less trivial to estimate. These can be determined by injecting a known energy \(E_{\mathrm{dep}}\) into a qubit island, and performing a 2D ML fit over those two variables for a stacked waveform. For a discussion of experimental tools that may enable such a calibration, interested readers are encouraged to read Ref.~\cite{MoffatMEMS,StifterMEMS}.

With \(\tau_{ss}\), \(r\), and \(\gamma\) determined using a set of calibration pulses, one can then assess the injected energy \(E_{\mathrm{dep}}\) of any other pulses (say, in a physics search period) as long as those parameters don't change significantly in time. This is done by performing a 1D maximum likelihood fit over \(E_{\mathrm{dep}}\) to determine reconstructed energy of the pulse. We call the resulting best-fit \(E_{\mathrm{dep}}\) the ``reconstructed'' energy \(E_{r}\). The energy determined in this way is manifestly non-negative, in contrast to energies that may be reconstructed according to the optimal filter prescription in Ref~\cite{GolwalaThesis}. 

\subsection{Estimating Energy Resolution and Threshold}
\label{subsec:SensorEnergyResolutionAndThreshold}

Energy resolution for this detection scheme is given by the width of a distribution of reconstructed event energies for a monoenergetic simulated in-qubit energy \(E_{\mathrm{dep}}\). In the following discussion, we simulate and reconstruct energies for populations of events with an instantaneous energy injection into a single qubit at an arbitrary time \(t=1\)~ms followed by time evolution of the system until \(t=6\)~ms. This time range is used for the maximum likelihood fits. The resulting energy resolution is dependent on a number of parameters:

\begin{figure}[t!]
\centering
\includegraphics[width=\linewidth]{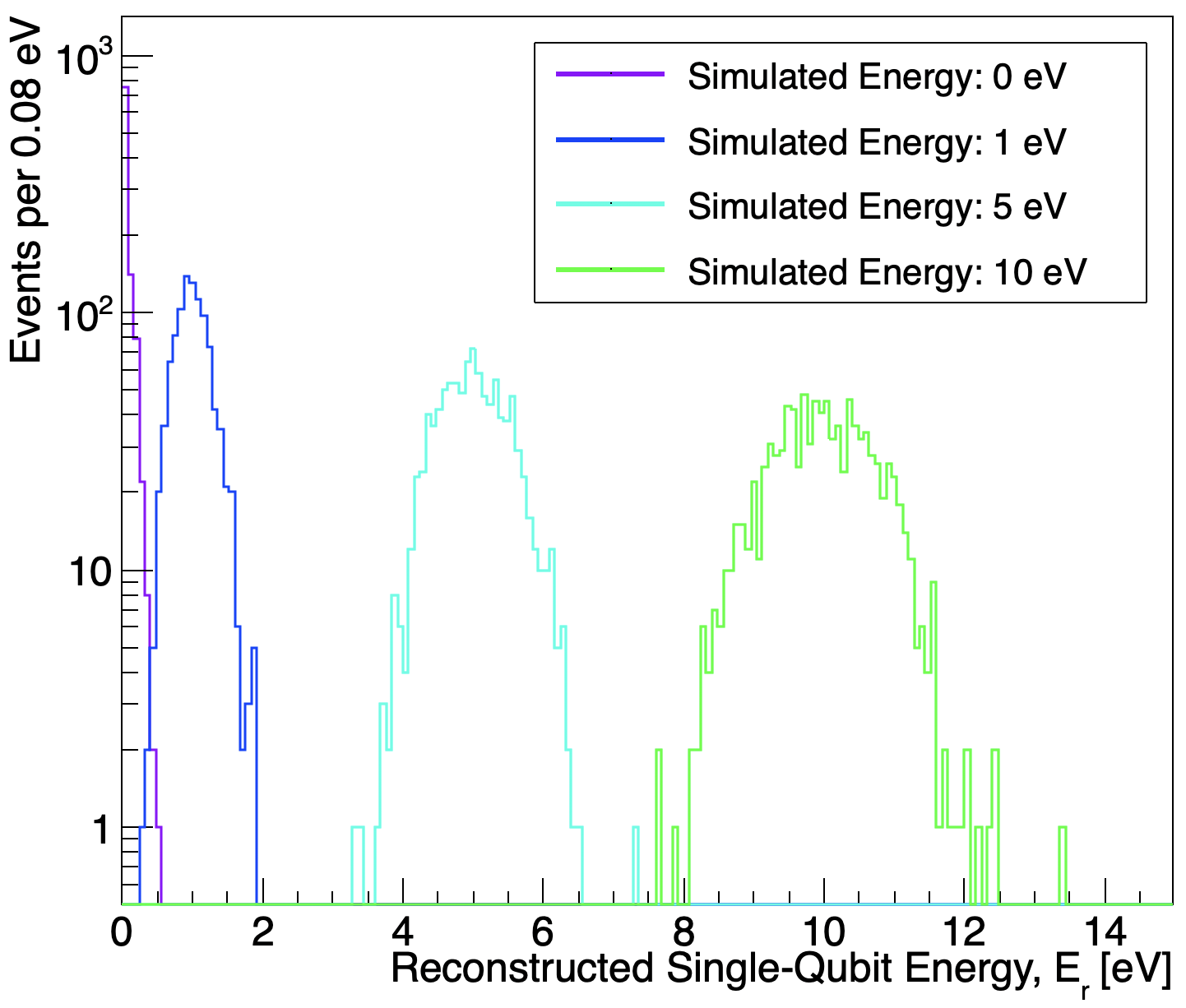}
\caption{Distributions of reconstructed energy \(E_{r}\) for simulated input energies \(E_{\mathrm{dep}}\) of 0~eV, 1~eV, 5~eV, and 10~eV injected into a single transmon island. Each distribution is formed from 1000 events run through QDR with baseline simulation parameters: \(T_{1,\mathrm{base}}=\)2~ms, \(\mathcal{F}=\)0.98.}
\label{fig:EnergyResolutionVsEnergy}
\end{figure} 

\begin{figure}[t!]
\centering
\includegraphics[width=\linewidth]{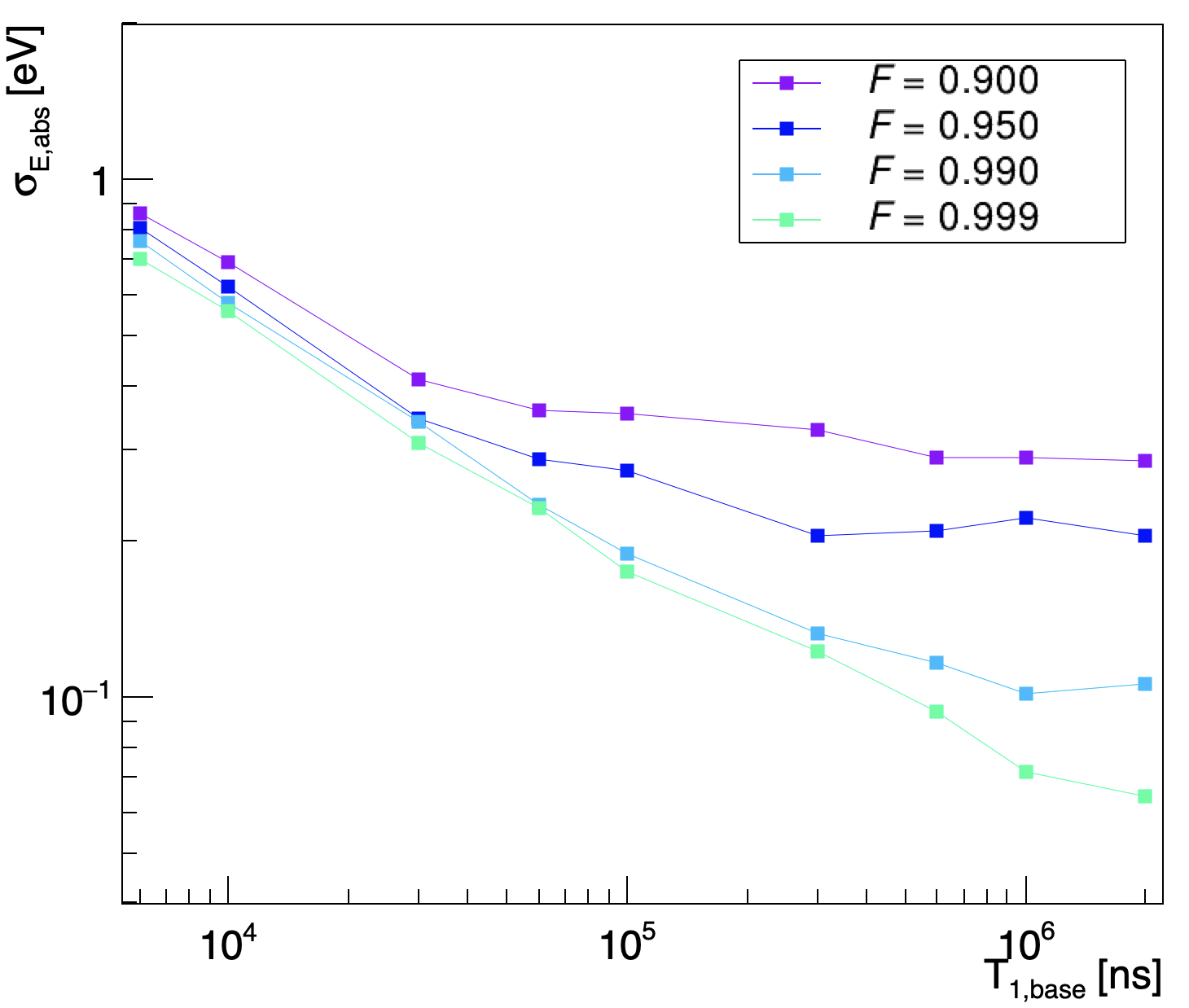}
\caption{Noise-only single-qubit sensor energy resolution \(\sigma_{E,\mathrm{abs}}\) for a search window of \(\Delta t=2~\upmu\)s and varying \(T_{1,\mathrm{base}}\) and \(\mathcal{F}\). Error bars indicating statistical uncertainty are included, and for most points are on the order of the size of the point itself.}
\label{fig:SigmaEabsVsT1AndSSF}
\end{figure}

\begin{itemize}
\item \(E_{\mathrm{dep}}:\) For a ``baseline'' near-term qubit performance characterized by \(T_{1,\mathrm{base}}=2\)~ms and \(\mathcal{F}=0.98\), and a \(\Delta t=2~\upmu\)s, Figure \ref{fig:EnergyResolutionVsEnergy} demonstrates the spread in reconstructed energy \(E_{r}\) as a function of injected \(E_{\mathrm{dep}}\). With increasing \(E_{\mathrm{dep}}\), the \(E_{r}\) distribution broadens considerably even from 1~eV to 10~eV. We attribute the broadening at these energies to the uncertainty (variance) associated with a binomial distribution defined by \(N\) measurements and success probability \(p_{\mathrm{obs}}\): \(Np_{\mathrm{obs}}(1-p_{\mathrm{obs}})\). For each waveform sample, this variance is largest for \(p_{\mathrm{obs}}=0.5\), leading to an increase in the overall uncertainty in \(E_{r}\) as we progress from near-threshold waveforms to those created by 5~eV or 10~eV energy depositions. However, even for \(E_{\mathrm{dep}}\) large enough that \(p_{\mathrm{obs}}\) is initially above 0.5 (which occurs \(\gtrsim\)20~eV), distributions do not re-narrow according to the binomial distribution variance. At these higher energies, a saturation in waveform shape sets in that makes energy reconstruction highly sensitive to small variations in waveform. Though rigorous exploration of this effect depends on details of phonon arrival times and is beyond the scope of this first model, we generally expect energy resolution to worsen monotonically with increased energy. Overall, we understand the nonlinear response of this readout scheme to be primarily attributed to the binomial nature of the signal read out, and will likely be present in other readout schemes where the signal is composed of discrete probabilistic measurements of a quantum state.

A notable quantity of interest is the ``noise-only'' energy resolution of a single qubit sensor, \(\sigma_{E,\mathrm{abs}}\), corresponding to the spread in the 0~eV curve in Figure \ref{fig:EnergyResolutionVsEnergy}. This is the resolution due to the jitter in a reconstructed energy that is fundamentally attributable to physical noise sources (from imperfect \(\mathcal{F}\) and finite \(T_{1,\mathrm{base}}\)), and is a metric one can use to compare the low-threshold performance of this detection scheme with low-threshold performance of other pairbreaking sensor architectures. Since energies are strictly non-negative, this curve has an appreciable sub-population of reconstructed events at exactly zero reconstructed energy. This distribution is therefore approximated well by a rectified Gaussian distribution derived from a pure Gaussian described by a mean of 0 and a \(\sigma=\sigma_{E,\mathrm{abs}}\). We may numerically estimate \(\sigma_{E,\mathrm{abs}}\) from this population by fitting the \(E_{r}>0\) region to a gaussian peaked at 0, and find for our baseline simulation that \(\sigma_{E,\mathrm{abs}}=0.088\)~eV. These fits are consistent with analytical estimates of the variance of the underlying rectified gaussian as given by Ref.~\cite{rectifiedGaussian}. Consistent with the convention used in Refs.~\cite{temples2024performance,WatkinsThesis}, we define the sensor energy threshold \(E_{\mathrm{thr}} \equiv 5\sigma_{E,\mathrm{abs}}\), which for our baseline simulation gives \(E_{\mathrm{thr}}=0.44\)~eV.
\item \(T_{1}\) and \(\mathcal{F}\): Both finite \(T_{1,\mathrm{base}}\) values and imperfect \(\mathcal{F}\) will ultimately contribute noise to this readout scheme. The effect of these parameters on \(\sigma_{E,\mathrm{abs}}\) is explored in Figure \ref{fig:SigmaEabsVsT1AndSSF}. Longer \(T_{1,\mathrm{base}}\) values and higher \(\mathcal{F}\) values are better, as they limit false positives from non-QP decoherence and readout error, respectively. Lower \(\mathcal{F}\) values will also lead to a higher leveling-off of \(\sigma_{E,\mathrm{abs}}\) with increasing \(T_{1,\mathrm{base}}\): for a fixed \(\mathcal{F}\), even an infinitely long coherence time will not continue to improve the overall signal-to-noise ratio. We provide additional analytical justification for this trend in Appendix~\ref{app:DependenciesOfSigmaEabs}. We also note that the assumptions underlying our QP time evolution will break down in the limit of small numbers of QPs produced by an event -- in such few-QP scenarios we cannot treat \(x_{qp}\) in the junction leads as being approximately equal to \(x_{qp}\) in the qubit island. For aluminum, this may cause deviation from the curves in Figure \ref{fig:SigmaEabsVsT1AndSSF} for in-qubit energies below approximately 100~meV in our Xmon design, though we note that this figure will vary depending on relative volumes of the qubit island and junction leads.

\item Search window \(\Delta t\): As long as \(\Delta t\ll T_{1,base}\), a longer \(\Delta t\) generally yields a lower \(\sigma_{E,\mathrm{abs}}\). While we explore this analytically in Appendix~\ref{app:DependenciesOfSigmaEabs}, the intuitive justification for this trend is that shorter \(\Delta t\) values lead to more frequent measurements, which, assuming an imperfect \(\mathcal{F}\), implies more opportunities for failed readout to induce errors that mimic signal pulses. In this regard, higher \(\mathcal{F}\) values will lead to a weaker dependence of \(\sigma_{E,\mathrm{abs}}\) on \(\Delta t\): when \(\mathcal{F}\) is unity, the noise-only energy resolution is flat in \(\Delta t\). Ultimately, a selection of a truly optimal \(\Delta t\) will in practice be constrained by other considerations such as a required degree of timing granularity (for example, needed to study the rising edge in a pulse). For example, with a near-term \(T_{1,base}=2\)~ms, it is impractical to use a \(\Delta t\) that is large enough that it violates \(\Delta t\ll T_{1,base}\): such waveforms would only enable a handful of individual relaxation measurements over the few-ms-long quasiparticle decay time, and would not give any detailed shape to a quasiparticle pulse. Additional considerations, including a discussion of how \(\Delta t\) impacts resolution and signal-to-noise for nonzero energy depositions, are explored in Appendix~\ref{app:DeltaTSNR}.
\end{itemize}

These estimates of \(\sigma_{E,\mathrm{abs}}\) are expected to be slightly better (lower) than practically achieved values, because our current simulations only model smearing from a limited number of effects. We do not account for temporal variations in our experimental quantities like \(T_{1,\mathrm{base}}\) or \(\mathcal{F}\), which will add additional spread to reconstructed energies beyond what's shown here. There will also be spread induced from variation in other parameters in Equation~\ref{eq:pr}. For example, if there is experimental noise in the qubit frequency (either due to flux noise in the case of split transmons or charge noise in the case of single-junction transmons), then this will add additional variation to the reconstructed energy around the baselines we have shown. These estimates are also likely to worsen slightly in future estimates as we add into QDR a residual occupation of the qubit excited state and the possibility for nonequilibrium QPs to cause upward energy transitions in the qubit, both of which are effects that have been demonstrated experimentally~\cite{ExcitedStateProbKulikov,ExcitedStateProbWenner}. Finally, we have made these estimates under the condition that the time of the initial in-qubit energy injection (and therefore the start time for the ML fits) is known. For a more realistic ``physics search'' scenario where the times of particle impacts are not known, energy resolution of reconstructed pulses may be larger than what we have quoted.

\begin{table}[t!]
\centering
\caption{A summary of the single-qubit sensor energy resolution for our baseline Xmon design (top) and our baseline design with collection fins and QP trap assumed.}
\begin{tabular}{|p{5.8cm}|p{1.8cm}|}
\hline
Simulation Quantity & Value \\
\hline
\hline
Input Parameter: \(T_{1,\mathrm{base}}\) & 2~ms\\
Input Parameter: \(\mathcal{F}\) & 0.98\\
Input Parameter: Search window \(\Delta t\) & 2~\(\upmu\)s\\
Performance (Xmon): \(\sigma_{E,\mathrm{abs}}\) & 0.088~eV \\
Performance (Xmon): \(E_{\mathrm{thr}}\) & 0.44~eV\\
Performance (collection-fins): \(\sigma_{E,\mathrm{abs}}\) & O(0.001)~eV \\
Performance (collection-fins): \(E_{\mathrm{thr}}\) & O(0.005)~eV \\
\hline
\end{tabular}
\label{table:SummaryTable}
\end{table}
\vspace{5mm}

Given these results, we comment on the single-qubit energy resolution and threshold realistically achievable for near-term detectors based on this readout scheme and demonstrated qubit architectures. Maximum published coherence times for 2D transmons are at the 300-400~\(\upmu\)s scale~\cite{WangMultiChip}. As a result, while an experimental achievement of our baseline simulated coherence time of \(T_{1,\mathrm{base}}=2\)~ms in a transmon qubit has yet to be published, we expect that such coherence may plausibly be achieved in the near-term. \(\mathcal{F}\) values of 0.995 have been experimentally demonstrated even without the use of a quantum-limited amplifier~\cite{ChenFidelity}, suggesting that our baseline \(\mathcal{F}\)=0.98 is realistic. Together, these observations suggest that our quoted single-qubit sensor energy resolution \(\sigma_{E,\mathrm{abs}}=0.088\)~eV and threshold \(E_{\mathrm{thr}}=0.44\)~eV are realistic targets in the near term with this sensing scheme. These figures are summarized in Table~\ref{table:SummaryTable}. For comparison, it is also useful to estimate \(\sigma_{E,\mathrm{abs}}\) for typical devices demonstrated in recent literature. Using the device in Ref.~\cite{MITRelaxation}, whose qubits have a \(\mathcal{F}\) of approximately 95\(\%\) and \(T_{1,\mathrm{base}}\) of around 40~\(\mu s\), Figure~\ref{fig:SigmaEabsVsT1AndSSF} suggests that \(\sigma_{E,\mathrm{abs}}\simeq0.3\)~eV and \(E_{\mathrm{thr}}\simeq1.5\)~eV for such ``currently achieved'' devices.

For further comparison, we also entertain coarse estimates of sensor thresholds for the chip designs employing collection fins. As these designs (which have an attached resonator) are based on the design in Ref.~\cite{SQUAT} which doesn't use a readout resonator, we necessarily must assume that a collection fin design with a resonator may be successfully operated in the transmon regime using the relaxation-style readout scheme. The increased collection area of this design implies an increase in (degradation of) \(\sigma_{E,\mathrm{abs}}\) through the increased volume \(V\). To counteract this, it is possible to design lower-\(\Delta\) QP trapping structures near the junction, as discussed in Ref~\cite{SQUAT}. This not only reduces the effective volume of the QPs near the junction to the much smaller trapping region volume, but also results in a QP gain process as near-gap QPs in the fins cascade to lower energy in the trap. From the formalism in Ref.~\cite{SQUAT} and similar design specifications, approximately 4\(\%\) of the QPs from a energy deposition in a collection fin can fall into the trap, be multiplied with a gain of \(\simeq\)7, and be localized within a volume O(0.02\(\%\)) that of a full collection fin. When propagated through Equation~\ref{eq:TimeIndependentDecoherence}, these effects combine to give a factor of O(100) increase in the decoherence rate for a given \(E_{\mathrm{dep}}\) relative to our baseline Xmon design and readout parameters. As a result, it may be possible to achieve sensor resolutions at the O(1)~meV scale and sensor thresholds at the O(5)~meV scale with such traps, though such estimates are likely idealized given the fact that these energy scales represent single- or few-phonon impacts. Moreover, these estimates rely on maintaining the same baseline readout performance (\(T_{1,\mathrm{base}}\), etc.), which may be significantly more challenging given the propensity of ambient QPs to diffuse into the trap region near the junction.


\section{Transmon Array Response to Energy Depositions}
\label{sec:ChipResponse}

Combining the phonon and single-qubit responses, one may now arrive at the response of the entire chip to energy depositions, and estimate an overall in-chip energy threshold. We reconstruct the full energy in the chip as 
\begin{equation}
\label{eq:ReconstructedEnergyInChip}
E_{r,\mathrm{chip}} = \frac{\sum_{i}E_{r,i}}{\eta_{ph,sp}}.
\end{equation}
Here, we use a sum over single-qubit reconstructed energies \(E_{r,i}\), rather than a fit to a summed waveform, to get the total in-chip reconstructed energy. As different qubits may have different values of \(r\) and \(\tau_{ss}\) depending on their fabrication, it is important to do individual calibrations and fits to each qubit waveform -- a single fit to a summed waveform may wash out the differences in these parameters and degrade energy reconstruction. While calculating \(E_{r,\mathrm{chip}}\), we continue to reconstruct \(E_{r,i}\) assuming an energy injection into a qubit at exactly the time of energy injection into the substrate. In reality, phonons will take 10-100~\(\upmu\)s to propagate around the chip before striking the sensor. This will introduce a slight misreconstruction of our \(E_{r,i}\), but that misreconstruction should be small as long as the characteristic quasiparticle fall time \(\tau_{ss}\) is much longer than that phonon diffusion timescale.

For a collection of events occurring spatially uniformly within the chip, Equation \ref{eq:ReconstructedEnergyInChip} implies that the mean \(E_{r,\mathrm{chip}}\) should match the true mean deposited in-chip energy. However, it will also introduce a spread in \(E_{r,\mathrm{chip}}\) that is dependent on the spatial nonuniformity of the phonon collection efficiency: \(E_{r,\mathrm{chip}}\) will be overestimated for events close to the qubits and underestimated for events far from the qubits. This can be mitigated somewhat if one uses qubit spatial information with the single-qubit signals to reconstruct a coarse position of the interaction within the qubit. Such a position reconstruction capability, while beyond the scope of this work, could enable a position-based correction of the reconstructed energy, reducing the spread in energies from a monoenergetic source even if the phonon absorption efficiency isn't perfectly uniform. Moreover, this position reconstruction capability might also enable sophisticated coincidence-based and position-based tagging of pathological interactions from nearby high-energy radioactivity and cosmic ray muons, which is invaluable in a low-threshold search.

\begin{figure}[t!]
\centering
\includegraphics[width=\linewidth]{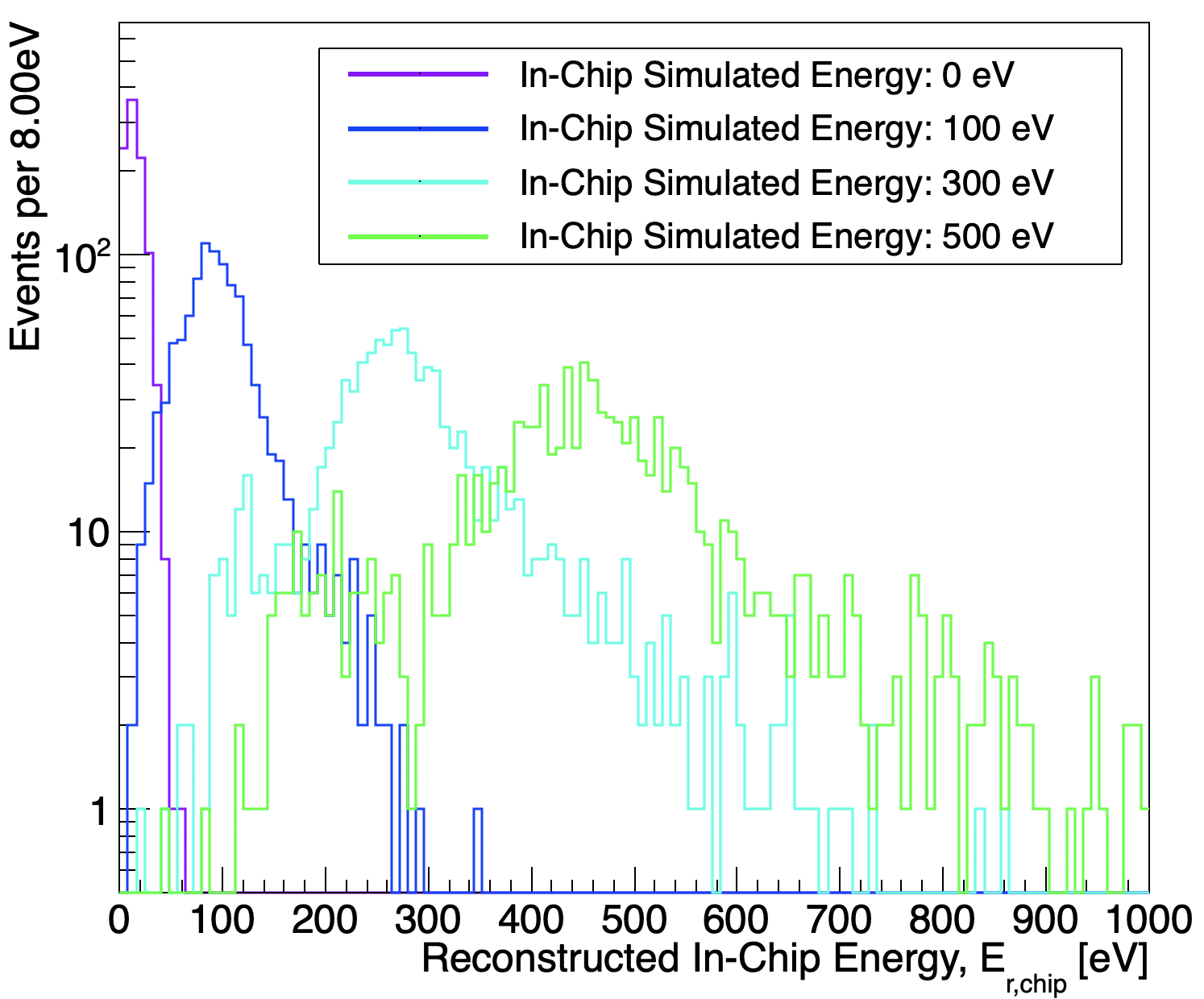}
\caption{Distributions of \(E_{r,\mathrm{chip}}\) for various injected energies, for chip Design 3 and our baseline single-qubit parameters. Here, the 0~eV, ``noise-only'' distribution is peaked at nonzero reconstructed energy. This is due to the explicitly non-negative reconstructed single-qubit energies \(E_{r,i}\) involved in the sum in Equation \ref{eq:ReconstructedEnergyInChip} (consistent with a distribution composed of a sum of several rectified Gaussian variables). The non-Gaussian shape of the 100, 300, and 500~eV cases is due to the nonuniformity in phonon collection efficiency across the chip.}
\label{fig:RecoEnergyInChipVsEnergy}
\end{figure}

With an in-chip reconstructed energy defined, we may proceed to study in-chip energy resolution and threshold. Figure \ref{fig:RecoEnergyInChipVsEnergy} shows several distributions of reconstructed energies for events simulated uniformly in XY in our baseline 6-qubit chip (Design 3 from Table~\ref{table:ConfigurationsVsThresholds}). For each event, a true total phonon energy was spawned from a random position in the chip as a sum of 20~meV phonons. This allows us to remain agnostic to the initial type of energy deposit but ignores a small degree of smearing in the initial phonon locations (like what would be expected from freed charge traveling a distance through the chip prior to phonon creation). As with the single-qubit case, the spread in each distribution grows with injected energy. The non-gaussianity of the histograms with nonzero simulated energy is due to the nonuniform phonon collection efficiency throughout the chip coupled with a uniformly drawn distribution of initial event locations throughout the chip.

To understand in-chip ``noise-only'' resolution and threshold, we make use of the 0~eV in-chip-energy curve in Figure~\ref{fig:RecoEnergyInChipVsEnergy}. The peak of this distribution is offset from \(E_{r,\mathrm{chip}}=0\), as expected for the sum of 6 random variables each described by identical rectified Gaussian distributions~\cite{rectifiedGaussian}. As a result, a noise-only energy resolution and threshold cannot be quoted in the exact same way as for the single-qubit case or for other pairbreaking detectors which display Gaussian noise profiles. However, a useful threshold for comparison to other detectors can still be quoted as the energy at which the noise-only distribution reaches the same statistical significance level as a 5\(\sigma\) deviation from the mean of a Gaussian. For a scenario in which each qubit's response is described by the same underlying rectified Gaussian distribution, this threshold is dependent on the number of qubits \(N_{q}\) and can be reasonably modeled as
\begin{equation}
\label{eq:InChipThresholdRGBased}
E_{\mathrm{thr,chip}} \simeq \frac{\sigma_{E,\mathrm{abs}}}{\eta_{ph,sp}}\Bigg[\frac{N_{q}}{\sqrt{2\pi}}+aN_{q}^{b}\Bigg].
\end{equation}
where \(a\simeq4.27\) and \(b\simeq0.44\). The first term in brackets represents the mean of the noise-only distribution, and can be derived using the moments of the rectified Gaussian distribution~\cite{rectifiedGaussian}. The second term captures the distance from that mean to the threshold. Its approximate power-law form is extracted computationally from Monte-Carlo-generated distributions corresponding to sums of \(N_{q}\) rectified Gaussian distributions, a process which also gives the above-quoted values for \(a\) and \(b\).

Table \ref{table:ConfigurationsVsThresholds} gives this \(E_{\mathrm{thr,chip}}\) value for the various chip parameters tested in Section \ref{sec:PhononPropagation}. We observe that the most optimal overall in-chip thresholds for our baseline Xmon hardware design are achieved using a limited ground plane, for which thresholds are as low as 41~eV. As this is still well above 1~eV, it is clear that achieving the goal of a sub-eV threshold will require additional iteration on qubit/resonator/control line design beyond the variations we have performed on the ``baseline'' Xmon architecture in Designs 1-12. Designs 13 and 14 illustrate a potential such iteration: if transmons with collection fins as in Ref.~\cite{SQUAT} can be achieved and read out in an energy relaxation readout scheme, the benefit of the larger collecting area and QP trap may enable sub-eV in-chip thresholds near the 100~meV scale.

Of some additional utility in thinking forward to larger-scale detector design is the fact that the full-chip ``simple'' threshold may also be approximated analytically. Using equation \ref{eq:InChipThresholdRGBased} and substituting in Equation~\ref{eq:EtaAnalyticalSpecific}, we arrive at 
\begin{multline}
\label{eq:InChipThresholdRGBasedFull}
E_{\mathrm{thr,chip}} = \sigma_{E,\mathrm{abs}}\Bigg[\frac{N_{q}}{\sqrt{2\pi}}+aN_{q}^{b}\Bigg]\times \\ \Bigg[\frac{N_{q}A_{t}p_{a,s}+A_{c}p_{a,c}+f(N_{q})}{N_{q}A_{q}p_{a,s}}\Bigg].
\end{multline}

\begin{figure}[t!]
\centering
\includegraphics[width=\linewidth]{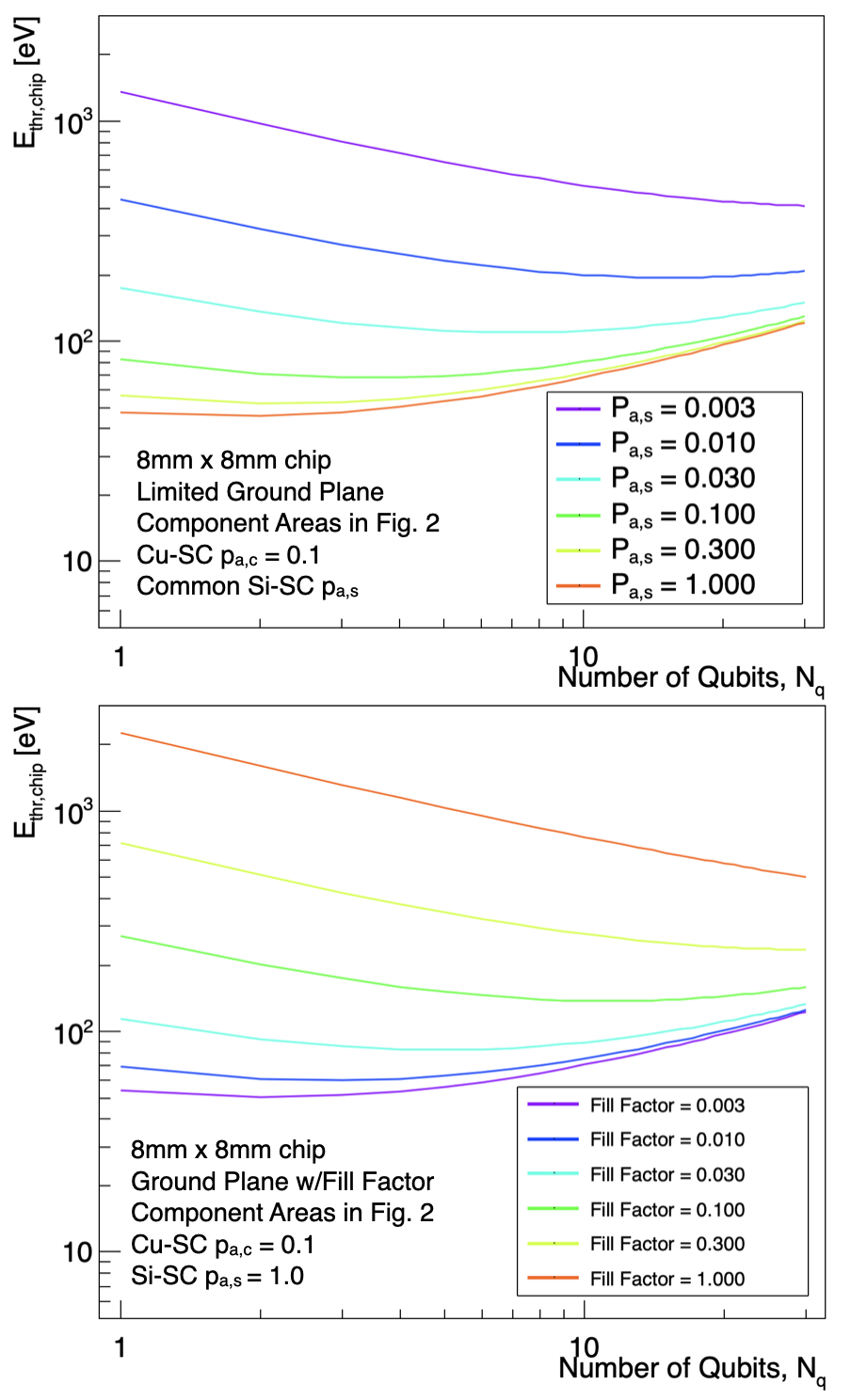}
\caption{Chip threshold vs. \(N_{q}\), with chip parameters described shown in the bottom left of each plot. In both scenarios we use the Xmon-style island design and assume \(\sigma_{E,\mathrm{abs}}=0.088\)eV. Top: variations in a common Si-SC \(p_{a,s}\) in a limited ground plane scenario. Bottom: variations in ground plane fill factor for \(p_{a,s}=1.0\).}
\label{fig:ChipThresholdVsNQubits}
\end{figure}

Figure~\ref{fig:ChipThresholdVsNQubits} shows results of this calculation for two sets of scenarios: one in which a common \(p_{a,s}\) is parameterized in a limited ground plane scenario, and one in which a ground plane with a parameterized fill factor is implemented.\footnote{This fill factor creates a scenario somewhere between our initially-stated ``Full'' and ``Limited'' ground plane scenarios.} For low absorption probabilities and/or high ground plane fill factors, the gain in phonon collection efficiency from adding qubits largely outweighs the additional noise for small numbers of qubits \(N_{q}\). However, at higher absorption probability (equivalently, lower fill factor), the benefit of gained phonon collection efficiency \(\eta_{ph,sp}\) balances the additional noise, showing an optimum \(N_{q}\) for minimizing chip threshold.\footnote{This optimum assumes no degradation of the individual sensor threshold upon adding more qubits to the chip. As qubit \(T_{1,base}\) values tend to drop with increasing \(N_{q}\) in a way that is commonly attributed to on-chip high-frequency environmental noise, this estimate may therefore overestimate the optimum number of qubits~\cite{WangMultiChip}.} As an example of applying this above expression to practical devices, we use it to calculate the phonon collection efficiency and \(E_{\mathrm{thr,chip}}\) for a ``currently-achieved'' chip, again using the device in Ref.~\cite{MITRelaxation} as our model. Given the transmon and chip dimensions shown in their work and our own assumption of \(p_{a,s}=1.0\) for the superconductor (which is not unreasonable due to their thick film), we estimate an \(\eta_{ph,sp}\sim0.7\%\). Using \(N_{q}=10\) and recalling our earlier estimate of \(\sigma_{E,\mathrm{abs}}\simeq0.3\)~eV, Equation~\ref{eq:InChipThresholdRGBased} gives \(E_{\mathrm{thr,chip}}\simeq675\)~eV.

Though it is often challenging to know the absorption parameters in Equation~\ref{eq:EtaAnalyticalSpecific} \textit{a priori}, this model for threshold may provide some intuitive guidance for how to select parameters of a future array of qubits for dark matter detection. In a scenario where a chip mass is fixed, we propose the following general strategy for creating a chip layout that has been optimized for achieving low in-chip energy thresholds. First, one chooses a minimum inter-qubit distance based on RF performance grounds -- qubits that are too closely packed can become capacitively coupled and may swap electromagnetic quanta, degrading isolated qubit performance~\cite{TransmonCouplings}. Next, one can use estimates of the areas of various phonon-absorbing regions on the chip \((A_{q},~A_{t}\), etc. from Equation~\ref{eq:EtaAnalyticalSpecific}) with estimates of the interfacial absorption probabilities \(p_{a,s}\) and \(p_{a,c}\) to determine an optimal number of qubits per chip. Lastly, those qubits are tiled on the chip at no closer than the minimum inter-qubit distance, and as spatially uniformly as possible so that \(\eta_{ph}\) is as uniform as possible throughout the chip. If more target mass is desired for exposure, additional chips may be added and read out in parallel. 

It is worth noting that a more realistic optimization may be driven by a budget-based constraint on the number of readout channels, with the less-expensive chip mass being allowed to vary to increase DM search exposure. While we do not attempt to solve this more general optimization problem here, the several rules of thumb we have discussed in the ``budget-unconstrained'' scenario (finding an optimal \(N_{q}\), uniform qubit spacing, etc.) will still help constrain this problem.


\section{Data-Model Comparison}
\label{sec:DataModelComparison}

We now demonstrate a coarse viability of our models and reconstruction by exercising them on a set of published experimental results that use the same energy relaxation readout scheme that we have studied in this work. In Ref.~\cite{MITRelaxation}, the effects of environmental radiation on a 10-transmon qubit chip (similar to Design 10 in Table~\ref{table:ConfigurationsVsThresholds}) were studied using waveforms constructed from a binned energy relaxation readout scheme like that described in Section~\ref{subsec:ReconstructionOfAnInQubitEnergyDeposition}. In this section, we demonstrate that for a reasonable set of assumed detector parameters, we can reconstruct a reasonable in-chip energy deposition for one of the events claimed to be from a cosmic ray muon passing through the chip.

The cosmic ray event shown in Figure~2 of Ref.~\cite{MITRelaxation} includes a waveform in each of 10 transmon qubits. Here, the relaxation readout scheme uses a \(\Delta t=3~\upmu\)s, an overall time-per-cycle of 15.3\(\upmu\)s, and a binning \(N\) of 40. Our goal is to fit these waveforms and extract estimated energies from them. To perform this fit, we need four additional ``calibration'' parameters: the \(\mathcal{F}\) and \(r\), \(\tau_{ss}\), and \(\gamma\) from Equation \ref{eq:pr}. The most trivial of these to estimate is \(\tau_{ss}\) -- this was already measured in Ref.~\cite{MITRelaxation} to be either 6~ms or 0.7~ms, depending on the junction of a given qubit. For this study, we only attempt to study those waveforms with \(\tau_{ss}=\)~6~ms, which corresponds to the JJ design that is more susceptible to QP tunneling out of the qubit island.

To estimate \(\mathcal{F}\), we can make our standard assumption of a symmetric fidelity and establish a lower bound on the value of \(\mathcal{F}\) using pre-pulse baseline errors. For Ref.~\cite{MITRelaxation}, we use this pre-pulse period to estimate a minimum \(\mathcal{F}\) at the 80\(\%\) scale. Since a non-infinite \(T_{1,\mathrm{base}}\) also realistically contributes to this error baseline, we assume for our calculations a symmetric \(\mathcal{F}\) of 95\(\%\), which can be estimated using the approximate scale of \(T_{1,\mathrm{base}}\) reported in Ref.~\cite{MITRelaxation} together with Equation~\ref{eq:FullSignal}. Using that \(\mathcal{F}\), taking \(E_{\mathrm{dep}}=0\), and measuring an average pre-pulse baseline error probability for a pulse, we can fully invert Equation~\ref{eq:FullSignal} for \(\gamma\).

The recombination constant \(r\) is the hardest quantity to estimate with the data provided. This difficulty is in large part due to uncertainty of the ``suppression factor'' \(F\) that should be divided into \(r\) to account for recombination phonons that re-break Cooper pairs before leaving the superconducting film. From Refs~\cite{Wang,KaplanAcousticMatching}, this suppression factor should be in the range of 5-10 for interfaces of Sapphire with 80-nm-thick Al film.\footnote{It is interesting to note that the measurements of this suppression factor mentioned in Ref.~\cite{KaplanAcousticMatching} seem to be somewhat inconsistent with the theoretical estimates of the phonon absorption length projected from Ref.~\cite{McEwen}. While in this case we assume the measured values for our approximation, it is worth mentioning that there is a substantial acknowledged degree of uncertainty in this parameter.} In the case of Ref.~\cite{MITRelaxation} where a Si chip with a 250~nm Al film is used, the change in substrate-film acoustic mismatch approximately cancels the thicker film, and as a result we assume in this calculation \(F\simeq10\).

\begin{figure}[t!]
\centering
\includegraphics[width=\linewidth]{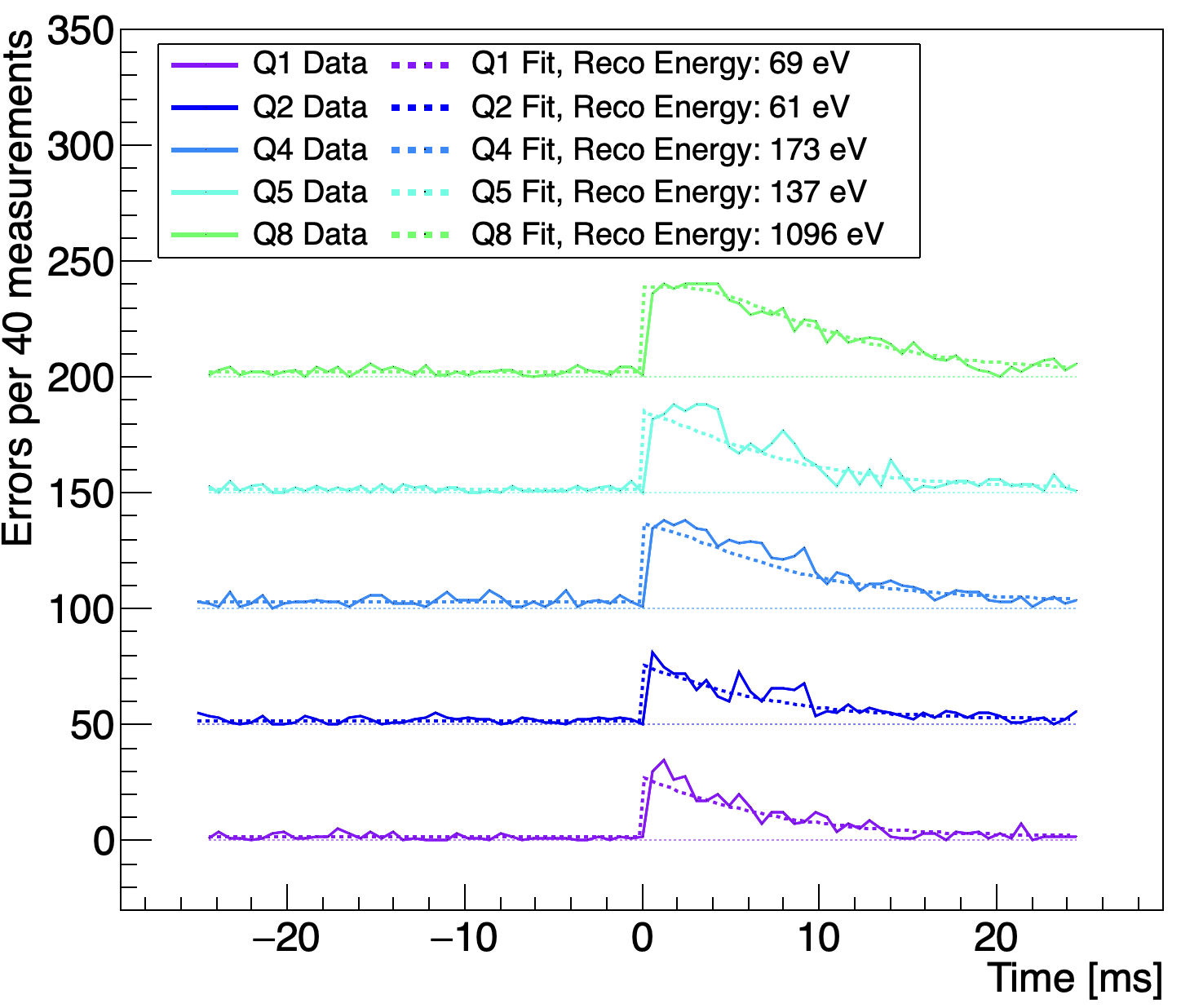}
\caption{Event waveforms digitized from Figure 2a of Ref.~\cite{MITRelaxation}, with fits employing the energy reconstruction technique developed in this work. Waveforms are offset vertically at intervals of 50 for clarity, with the ``zero-point'' of each plot shown as a dotted line. Fit functions are shown as dashed lines. Reconstructed best-fit energies are included in the legend.}
\label{fig:MITRelaxationFits}
\end{figure}

Figure~\ref{fig:MITRelaxationFits} shows reproduced waveforms for Qubits 1, 2, 4, 5, and 8 in the 10-transmon chip, along with maximum-likelihood fits to the overall in-qubit energies. Waveforms were extracted by digitizing the plots in Ref.~\cite{MITRelaxation}. The estimated summed energy within all 5 qubits is 1.537~keV. If we compute a per-qubit average from these, each qubit has absorbed on average 307~eV in this event. 

Given the qubit dimensions and overall chip dimensions quoted in Ref.~\cite{MITRelaxation}, we can continue and estimate an in-chip overall energy. Using the estimated \(\eta_{ph,sp}\sim0.7\%\) together with our average in-qubit energy, we estimate a total in-chip energy of about 440~keV. For a minimum-ionizing particle traversing a Si chip of thickness 0.35~mm, this estimated energy deposition implies an incident angle of about 14\(\degree\) away from the plane of the chip~\cite{PDG}. Since the zenith is approximately in the plane of the chip according to Ref.~\cite{MITRelaxation}, this suggests that the muon also arrived at approximately 14\(\degree\) from the zenith. Given the displayed chip orientation and positioning relative to the other scintillation detectors, an incident particle track angled at 14\(\degree\) with respect to the zenith is reasonably consistent with the observation of a qubit-scintillator coincidence in this event and the conclusion that the event was caused by a cosmic muon.

Though we have deliberately averaged out position information for simplicity, this exercise also demonstrates that reconstruction of position information in an interaction is realistic using our energy reconstruction formalism. In Figure~\ref{fig:MITRelaxationFits}, Qubit 8 clearly has a more saturated waveform than Qubit 1, indicating that the bulk of the phonon energy emitted occurred closer to Qubit 8 (i.e. on the right side of the chip as presented in Ref~\cite{MITRelaxation}). This suggests that even at lower energies than this muon event, position information may indeed be applied to remove pathological background events and perform corrections on reconstructed energy to improve in-chip energy resolution.

Sources of uncertainty in this exercise include slight errors in plot digitization, the aforementioned worsening energy resolution for in-qubit energy depositions above \(\gtrsim20\)~eV, our assumptions about phonon absorption probability \(p_{a,s}\) at the Si-SC interface, our averaging of the qubits' energy depositions (ignoring spatial variation), our reconstruction assuming negligible phonon propagation time from the energy deposition to the qubit islands, and our previously-addressed assumptions about the recombination constant \(r\).\footnote{After writing, we became aware that the waveforms in Ref.~\cite{MITRelaxation} included a post-processing selection to compensate for the authors’ decision not to use an active reset in each cycle. As a result, the energies reconstructed from these waveforms are systematically underestimated. QDR simulations show that this underestimation is at the scale of a factor of 2. Even with this underestimation, we find the approximate reconstruction of these waveforms to be a good validation of our methodology. Future work will explore energy reconstruction with proper accounting of this post-processing.} However, even though these many uncertainties are challenging to characterize given the limited information available, we are able to construct a picture of this event as having an in-chip deposited energy consistent with a cosmic muon passing through the chip, in agreement with the conclusions of Ref.~\cite{MITRelaxation} claiming that the event's origin is a cosmic muon. This ability is a promising first step for better understanding the tools and formalism for using this superconducting qubit technology for detection purposes.


\section{Discussion}
\label{sec:Discussion}

Given the goal of using superconducting qubits as particle detectors, we must discuss how the results of the last three sections fit into the larger landscape of low-mass dark matter searches. For a near-term ``baseline'' chip design with Xmon qubits, our estimated chip energy threshold is 41~eV for demonstrated qubit architectures (though for yet-to-be-demonstrated ground plane designs), which is only a factor of a few higher than state-of-the-art TES-based detectors~\cite{RenQET} but competitive with state-of-the-art MKID detectors~\cite{temples2024performance}. To compete with these technologies, additional iteration on our baseline design is needed. There are three plausible avenues for improving detector performance: readout scheme, materials choices, and layout optimization.

First, the readout scheme we use plays a role in limiting our projected threshold. Here, the link between \(x_{qp}\) and \(\Gamma_{qp}\) in Equation \ref{eq:TimeIndependentDecoherence} ultimately determines our quoted sensor energy resolution \(\sigma_{E,abs}\). However, different readout schemes, such as those enabling mapping of the qubit island charge parity to the readout value, may offer increased sensitivity for nearly identical hardware. If such readout schemes are also composed of gate operations, they will also inherit much of the mathematical formalism discussed and many of the trends we have observed for the energy relaxation scheme presented here. Studying alternative readout schemes will be a focus of follow-up studies.

Second, materials choices may also have a major impact on the overall in-chip threshold through both \(\sigma_{E,abs}\) and \(\eta_{ph,sp}\). While our studies assume purely aluminum qubits, use of a lower-gap material may yield at least a \(1/\sqrt{\Delta}\) improvement of \(\Gamma_{qp}\) in Equation~\ref{eq:TimeIndependentDecoherence}~\cite{Dutta_2022}. Lower-gap materials such as Hf, AlMn, or IrPt are still an area of exploration within the qubit community, but they are regularly exploited by the TES and MKID communities for lowering threshold~\cite{IrPtTES,HfMKIDs,AlMnTESs}. The advantage of exploiting such lower gap materials is also evident in the coarsely estimated thresholds for the collection-fins Designs 13 and 14 if a QP trap is assumed to be present as in Ref.~\cite{SQUAT}. Moreover, materials design may also improve our phonon collection efficiency \(\eta_{ph,sp}\): if significant areas of on-chip superconducting structures are strictly needed, say, for RF performance (i.e. ground plane, transmission line, etc.), it is advantageous to make these structures from higher-gap material such as Nb (2\(\Delta\simeq\)3~meV)~\cite{NiobiumGap}. Here, the few-meV phonons from the in-substrate downconversion cascade are much closer to the superconducting gap, and by Equation \ref{eq:paVsEph} are therefore much more likely to stream back out of the film without breaking CPs. This enables better phonon collection in the target (Al) structures sensitive to QPs.

Finally, iteration on chip layout to reduce the phonon-insensitive on-chip superconductor is also a strategy for improving \(\eta_{ph,sp}\) and lowering in-chip threshold. While we nominally reach up to 17\(\%\) with Designs 13 and 14, there is a clear benefit to further studying methods by which to reduce the area of the readout resonators, transmission line, and qubit control lines. Some potential strategies for increasing the overall sensitive fraction include minimizing the total number of transmission lines through frequency multiplexing, removing the readout resonator~\cite{SQUAT}, and/or coupling multiple qubits to each resonator~\cite{EntanglementMetrology}. While thoroughly exploring the details of such additional designs is beyond the scope of this paper, it will likely be required to compete with leading TES-based technologies.


It is also instructive to consider impacts of this study on design of qubit chips for the quantum computing community. Correlated errors are undesirable as they impair quantum error correction (QEC) with the surface code, and as a result one generally wants to invert many of the decisions we have discussed for how to design a low-threshold detector. In particular, in-qubit phonon collection efficiencies should be minimized through use of a full ground plane and, if possible, high-\(\Delta\) qubit superconductor material and normal metal baths for phonon downconversion~\cite{Downconversion}. Junction gap engineering strategies can further suppress errors by restricting the diffusion of QPs out of the qubit islands, as demonstrated in Refs.~\cite{MITRelaxation,mcewen2024resisting}. Beyond this, a detailed optimization of the chip layout for minimizing correlated errors depends on the specific algorithm and physical layout of the chip, and is a topic for future study with G4CMP.




\section{Conclusions}
\label{sec:Conclusions}

In this work, we have performed an end-to-end modeling of the physics relevant to understanding the energy threshold of a detector based on superconducting qubits operated in a relaxation-detection readout scheme. We have explored the properties of phonon propagation from an interaction site to a qubit sensor using G4CMP for a variety of different chip design parameter values, and have estimated the corresponding phonon collection efficiencies \(\eta_{ph,sp}\). We've also explored the single-qubit response to phonon energy depositions using a custom-built simulation software called Quantum Device Response (QDR), and found that for an example energy relaxation readout scheme using Xmon qubits, we estimate a qubit sensor energy threshold of 0.44~eV for near-term qubit readout parameters. We then studied the combined effect of phonon transport and single-qubit readout to estimate thresholds of in-chip energy deposited, in which we found a 41~eV in-chip threshold if one can significantly limit the ground plane present in the chip. We also found that if a qubit design with collection fins and QP traps near the junction can still be operated with this readout scheme, the in-chip threshold may plausibly be lowered to the O(0.1~eV) scale. Finally, we have also demonstrated the viability of the formalism that we've used to develop these estimates via comparison to a real cosmic ray muon event in Ref.~\cite{MITRelaxation}, though further validation is needed.

While this estimated in-chip 41~eV threshold is still higher than other leading sensor technologies, it sets a benchmark from which qubit-based detector technology can be further improved. Moreover, the work laid out here may act as a guide for the development of other low-threshold qubit-based detectors: much of the formalism in Section~\ref{sec:SingleQubitResponse} may also be applied to more sensitive gate-based readout schemes, and the strategies in Section \ref{sec:ChipResponse} may enable coarse optimization of chip parameters given any new qubit/resonator/control line design and some knowledge of phonon absorption probabilities. Together, these strategies lay the groundwork for designing and understanding the sensitivity of superconducting-qubit-based detectors with gate-based readout schemes in the field of sub-GeV/c\(^{2}\) dark matter direct detection.


\begin{acknowledgments}
The authors would like to thank Patrick Harrington, Kyle Serniak, Joseph Formaggio, William Oliver, and the rest of the MIT Engineering Quantum Systems group for for significant correspondence on the application of our energy reconstruction technique to the data they presented in Ref.~\cite{MITRelaxation}. The authors would also like to thank Karthik Ramanathan, Sunil Golwala, Osmond Wen, Noah Kurinsky, and Mike Kelsey for useful discussions of phonon and quasiparticle microphysics in superconducting devices. We also thank Robert McDermott, Sohair Abdullah, and Sami Lewis for helpful intuition-building lessons about design and operation of transmon qubits. Finally, we thank Edgard Bonilla for insightful discussions about statistical distributions and tools for constructing and analyzing the energy relaxation signals. This manuscript has been authored by Fermi Research Alliance, LLC under Contract No. DE-AC02-07CH11359 with the U.S. Department of Energy, Office of Science, Office of High Energy Physics. This work was supported by the U.S. Department of Energy, Office of Science, National Quantum Information Science Research Centers, Quantum Science Center and the U.S. Department of Energy, Office of Science, High-Energy Physics Program Office.

RL led the organization and development of the simulations and analysis chain used in this work, and drafted the manuscript. RL, DBa, LH, EFF, and RK motivated the overall study and provided guidance in the scoping and presentation of this work. IH performed G4CMP phonon transport simulations useful in refining insights made in Section \ref{sec:PhononPropagation}. DT, SD, and KA contributed to development of the QDR package. DT also provided insight into community standards for energy resolution and calibration that motivated additional shaping of paper narrative. DBa, LH, EFF, RK, IH, KA, DBo, GB, GC, AC, RG, KS, and SS provided feedback and thoughtful discussions throughout the development of this work.
\end{acknowledgments}


\appendix

\section{Modeling Dependencies of \(\sigma_{E,abs}\)}
\label{app:DependenciesOfSigmaEabs}

As there are several complex steps involved in estimating the noise-only energy resolution \(\sigma_{E,\mathrm{abs}}\) via the QuantumDeviceResponse package and our maximum-likelihood energy reconstruction technique, we would also like to include an intuitive approach for understanding how this resolution scales with various parameters. We do this by claiming that this energy resolution must reasonably scale as the uncertainty in the number of errors \(\sigma_{\mathrm{e}}\) within a generic time window \(\delta t\) which contains a number \(N\sim\delta t/\Delta t\) of individual readout cycles. This dependence can be analytically expressed using binomial stastistics:
\begin{align}
\label{eq:AppA_1}
\sigma_{\mathrm{e}} & = \sqrt{Np_{obs}(1-p_{obs})} \\
& = \sqrt{\frac{\delta t}{\Delta t}p_{obs}(1-p_{obs})}
\end{align}
As we're studying the noise-only energy resolution, we construct \(p_{obs}\) using Equation~\ref{eq:FullSignal} with a true relaxation probability \(p_{r}=1-e^{-\gamma}\). Since \(\gamma=\Delta t/T_{1,\mathrm{base}}\) and since we will assume \(\Delta t \ll T_{1,\mathrm{base}}\), this allows us to rewrite \(\sigma_{\mathrm{e}}\) as:
\begin{equation}
\label{eq:AppA_3}
\sigma_{\mathrm{e}} = \sqrt{\frac{\delta t}{\Delta t}\Bigg[\frac{1}{2}+\mathcal{F}\Big[\frac{\Delta t}{T_{1,\mathrm{base}}}-\frac{1}{2}\Big]\Bigg]\Bigg[\frac{1}{2}- \mathcal{F}\Big[\frac{\Delta t}{T_{1,\mathrm{base}}}-\frac{1}{2}\Big]\Bigg]}.
\end{equation}
\begin{figure}[b!]
\centering
\includegraphics[width=\linewidth]{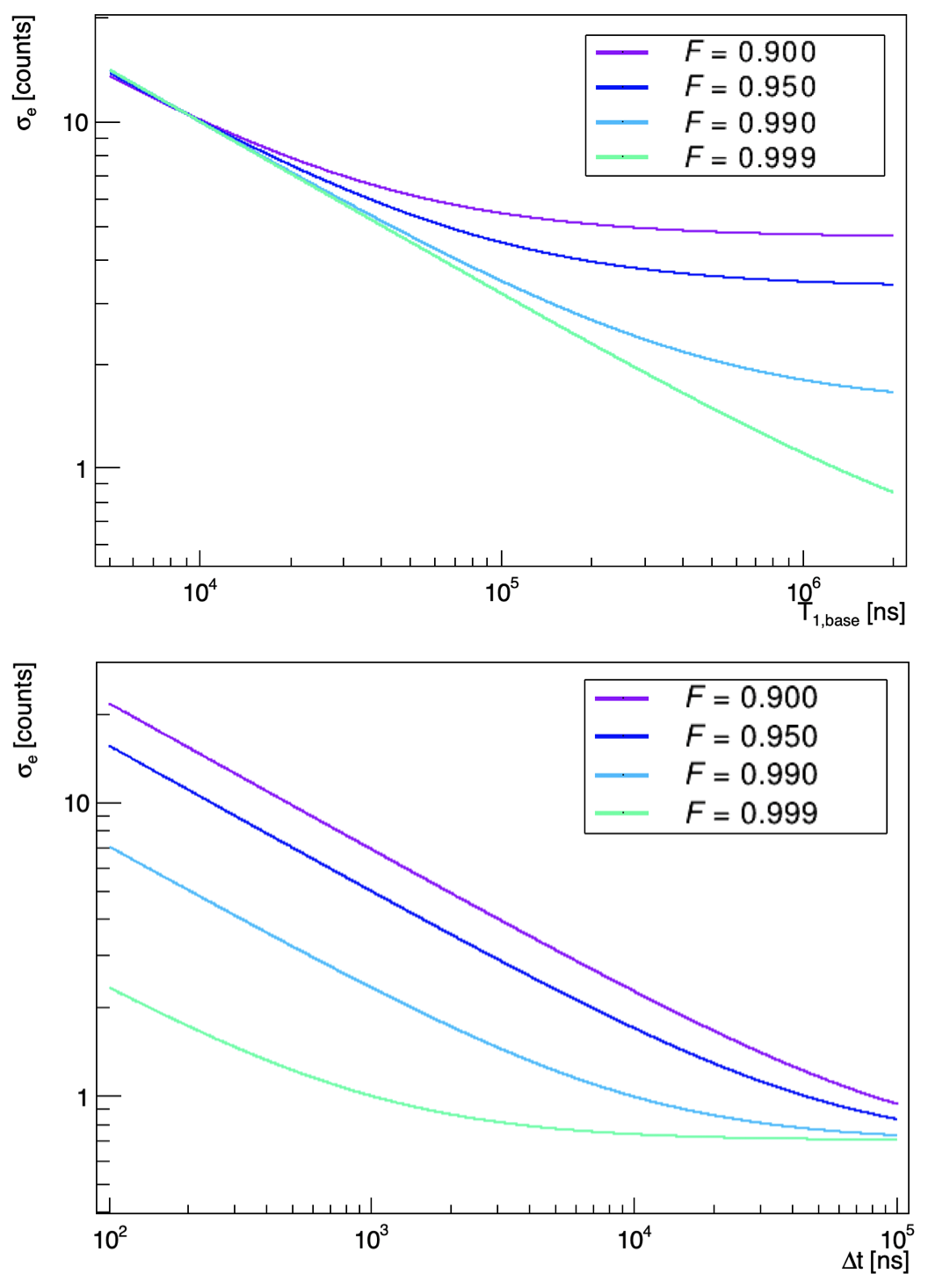}
\caption{Top: Scaling of \(\sigma_{\mathrm{e}}\) with \(T_{1,\mathrm{base}}\), for a \(\Delta t=2~\mu\)s. The intersection point at the left is a result of the approximate form used from Equation~\ref{eq:AppA_5}, which breaks down for low \(T_{1,\mathrm{base}}\). Bottom: Scaling of \(\sigma_{\mathrm{e}}\) with \(\Delta t\), for a \(T_{1,\mathrm{base}}=2\)~ms. In both cases, we arbitrarily use \(\delta t=1\)~ms, but this only affects the absolute scale of the vertical axis, and not the more important scaling with \(\Delta t\) or \(T_{1,\mathrm{base}}\).}
\label{fig:ModelForSigmaEAbs}
\end{figure} 

This can be further reduced to three terms:
\begin{align}
\label{eq:AppA_4}
\sigma_{\mathrm{e}} & = \sqrt{\frac{\delta t}{\Delta t}\Bigg(\frac{1}{4} - \mathcal{F}^{2}\Big(\frac{1}{4}-\frac{\Delta t}{T_{1,\mathrm{base}}}+\frac{\Delta t^{2}}{T_{1,\mathrm{base}}^{2}}\Big)\Bigg)}\\
\label{eq:AppA_5}
&\simeq \sqrt{\frac{\delta t}{\Delta t}\Bigg(\frac{1}{4} - \frac{\mathcal{F}^{2}}{4} + \frac{\mathcal{F}^{2}\Delta t}{T_{1,\mathrm{base}}}\Bigg)}
\end{align}
where in Equation~\ref{eq:AppA_5} we have again invoked \(\Delta t \ll T_{1,\mathrm{base}}\).

From this expression, we may understand the scaling of \(\sigma_{e}\) with fidelity, \(T_{1,\mathrm{base}}\), and \(\Delta t\). The top panel of Figure~\ref{fig:ModelForSigmaEAbs} shows the scaling of \(\sigma_{\mathrm{e}}\) with \(\mathcal{F}\) and \(T_{1,\mathrm{base}}\) for the same set of conditions shown in Figure~\ref{fig:SigmaEabsVsT1AndSSF}. From this we recover the same behavior discussed in Section~\ref{subsec:SensorEnergyResolutionAndThreshold}: for high \(\mathcal{F}\), \(\sigma_{\mathrm{e}}\) scales with the inverse square root of \(T_{1,\mathrm{base}}\). The scaling with \(\Delta t\), shown in the lower panel of Figure~\ref{fig:ModelForSigmaEAbs}, also follows: for \(\mathcal{F}=1\), Equation~\ref{eq:AppA_5} simplifies to
\begin{equation}
\label{eq:AppA_6}
\sigma_{e} = \sqrt{\frac{2\delta t}{T_{1,\mathrm{base}}}},
\end{equation}
which is flat in \(\Delta t\). In contrast, if \(\mathcal{F}<1\), then there are terms in Equation~\ref{eq:AppA_5} that give an inverse-root scaling with \(\Delta t\). When \(\Delta t/T_{1,\mathrm{base}}\ll 1\), these terms dominate the resolution and yield an increase in \(\sigma_{\mathrm{e}}\) with decreasing search window duration \(\Delta t\). As this is the same scaling we observed through use of QDR and our energy reconstruction technique, we have proposed this intuitive explanation for the scaling.

\section{Effect of Search Window \(\Delta t\) on Energy Resolution for Nonzero \(E_{\mathrm{dep}}\)}
\label{app:DeltaTSNR}

\begin{table*}[t!]
\centering
\caption{Joint probabilities of having either an error (left column) or no error (right column) and one of the four possible underlying physical processes occur during a single measurement cycle. Definitions of the quantities \(\eta\) and \(R\) can be found in text of Appendix~\ref{app:DeltaTSNR}.}
\begin{tabular}{|p{3.5cm}||p{4.5cm}|p{4.5cm}|}
\hline
Underlying Cause & Error & No Error \\
\hline
\hline
Signal-induced & \((1+\mathcal{F})(1-e^{-\eta R})e^{-\eta}/2\) & \((1-\mathcal{F})(1-e^{-\eta R})e^{-\eta}/2\) \\
Background-induced & \((1+\mathcal{F})(1-e^{-\eta})e^{-\eta R}/2\) & \((1-\mathcal{F})(1-e^{-\eta})e^{-\eta R}/2\) \\
Signal + Background & \((1+\mathcal{F})(1-e^{-\eta R})(1-e^{-\eta})/2\) & \((1-\mathcal{F})(1-e^{-\eta R})(1-e^{-\eta})/2\) \\
Fidelity-induced & \((1-\mathcal{F})e^{-\eta (R+1)}/2\) & \((1+\mathcal{F})e^{-\eta (R+1)}/2\) \\
\hline
\end{tabular}
\label{table:JointProbabilities}
\end{table*}
\vspace{5mm}

We now present a more rigorous discussion of how the search window duration \(\Delta t\) between the \(\pi\)-pulse and the measurement pulse affects the signal-to-noise in a given readout cycle, with an emphasis on scenarios where a nonzero \(E_{\mathrm{dep}}\) occurs. To do this, it is instructive to explore the various contributors to a measured error. We do this by considering four different scenarios: 
\begin{enumerate}
\item \textbf{``Signal-induced'' Error:} In this scenario, a quasiparticle signal of interest (say, from a scatter in the chip) triggers a relaxation, and the measurement successfully reads out an error.
\item \textbf{``Background-induced'' Error:} In this scenario, a relaxation occurs because of physical phenomena (QP poisoning, Purcell decay, etc.), and the measurement successfully reads out an error. In the same time, no relaxations from signal-induced QPs occur.
\item \textbf{``Signal + Background'' Error:} In this scenario, both signal-induced QP tunneling and background phenomena occur within the search window, which collectively produce a single relaxation that is successfully read out as an error.
\item \textbf{``Fidelity-induced'' Error:} In this scenario, no true relaxation occurs, but an error is read out due to poor single shot fidelity.
\end{enumerate}
These four broad options span the space of possibilities for physical phenomena preceding an error. For characteristic timescales \(\tau_{s}\) and \(\tau_{b}\) associated with signal- and background-induced relaxations respectively, an overall search window duration \(\Delta t\), and dimensionless variables \(\eta\equiv\Delta t/\tau_{b}\) and \(R\equiv\tau_{b}/\tau_{s}\), these outcomes have joint probabilities as shown in the left column of Table~\ref{table:JointProbabilities}. We also present their counterparts for a ``no-error'' readout in the same table. The corresponding probabilities in the table give the joint probabilities of having an error (or no error) coincident with the various  underlying physical scenarios.

The entries in Table~\ref{table:JointProbabilities} may also be used to compute the conditional probability of having an underlying physical scenario occur given a measured outcome. This allows us to define a signal-to-noise metric, which we take to be the probability \(p(s|e)\) that, given a measured error, the underlying physical scenario was a signal-induced relaxation. The expression for this follows from Bayes' theorem:
\begin{equation}
\label{eq:Bayes1}
p(s|e) = \frac{p(e|s)p(s)}{p(e)} = \frac{p(e,s)}{p(e)}.
\end{equation}

The denominator of this is just the sum over the joint probabilities in the left column in Table~\ref{table:JointProbabilities}. Using those expressions, this conditional probability becomes
\begin{equation}
\label{eq:Bayes2}
p(s|e) = \frac{(1+\mathcal{F})(1-e^{-\eta R})e^{-\eta}}{1+\mathcal{F}(1-2e^{-\eta (R+1)})}.
\end{equation}
Additional conditional probabilities for the other three underlying physical scenarios being coincident with a measured error can be calculted in an analogous way.

\begin{figure}[t!]
\centering
\includegraphics[width=\linewidth]{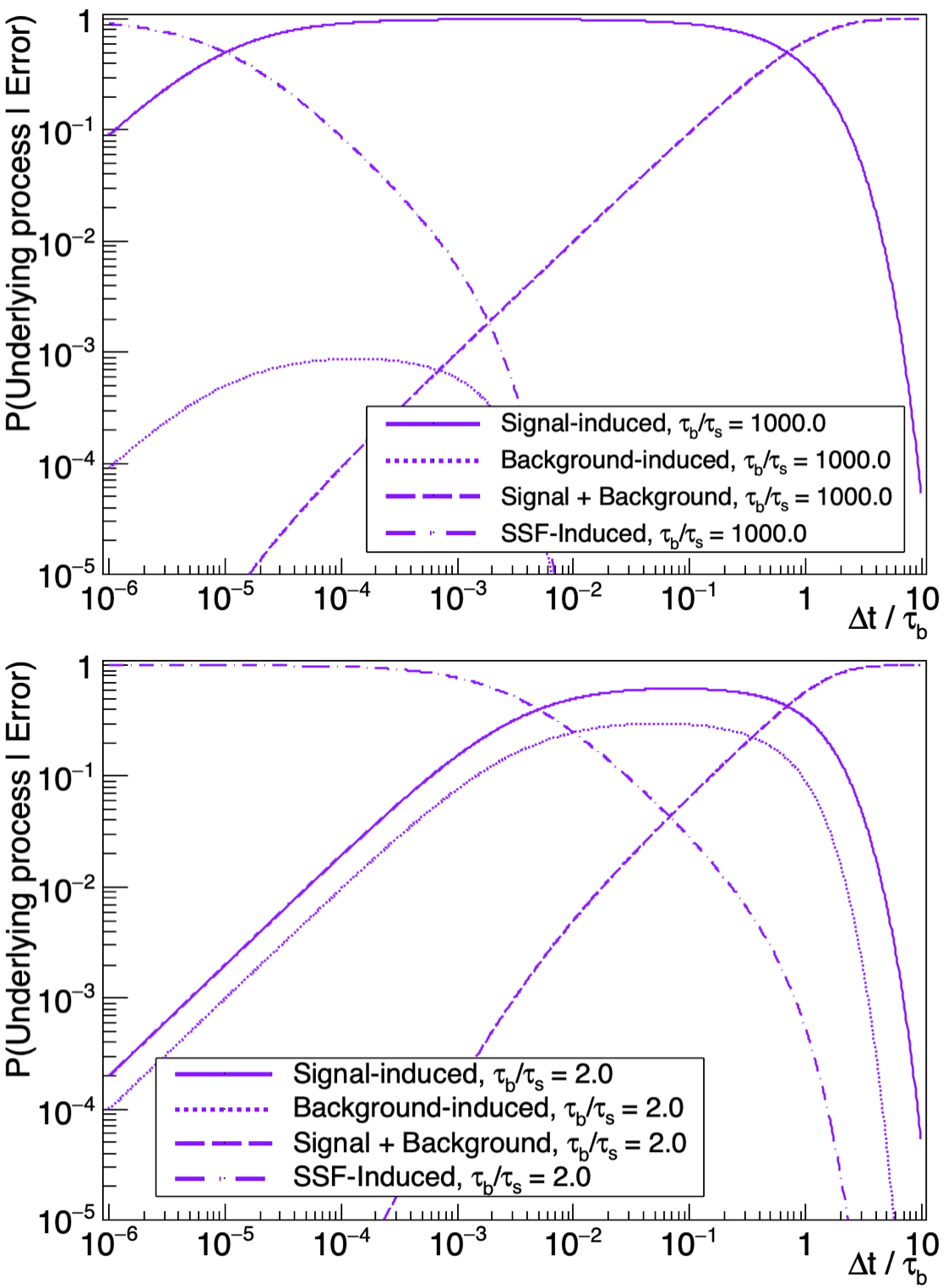}
\caption{Conditional probabilities of the four underlying processes in Table~\ref{table:JointProbabilities}, as a function of \(\eta\equiv\Delta t/\tau_{b}\), and for \(\tau_{b}/\tau_{s}\) of 1000.0 (top) and 2.0 (bottom). These are performed for a single-shot fidelity of 0.98.}
\label{fig:SNRPlotsDeltaT}
\end{figure} 

With the conditional probabilities calculated, we may explore the variation in our signal-to-noise by searching for where \(p(s|e)\) is maximized relative to the conditional probabilities representing background-induced errors \(p(b|e)\), signal+background-induced errors \(p(s+b|e)\), and fidelity-induced errors \(p(\mathcal{F}|e)\). While in principle a maximum to this curve can be found by differentiating Equation~\ref{eq:Bayes2}, we find it more useful to present and discuss a visual representation of the various conditional probabilities as shown in Figure~\ref{fig:SNRPlotsDeltaT}. 

There are three features of this plot worth pointing out. First, at the lowest \(\Delta t/\tau_{b}\), the probability that an error is fidelity-induced is dominant, as not enough time has elapsed in the search window to trigger a relaxation from either signal or background.\footnote{This development ignores additional relaxation during the measurement pulse, which is an additional complicating factor mentioned in the text but one we omit here for clarity of this development.} Second, at very high \(\Delta t/\tau_{b}\), the signal-to-noise metric \(p(s|e)\) falls steeply because (at least for \(\tau_{s}\lesssim\tau_{b}\)), the window is much longer than any characteristic relaxation time, and therefore it becomes challenging to unambiguously ascribe a signal-like nature to an error. Finally, the \(\Delta t\) of strongest signal-to-noise, the maximum of \(p(s|e)\), is found to be decrease with the ratio of signal strength to background strength, \(\tau_{b}/\tau_{s}=\Gamma_{s}/\Gamma_{b}\). This is unsurprising: for a given background rate and an increasing signal rate, a correspondingly decreasing \(\Delta t\) reduces the probability of S+B-induced backgrounds while keeping the signal-induced background probability steady. 

We have presented these estimates as a way of thinking about the signal-to-noise in a single relaxation measurement, but it is important to briefly comment on how these lessons extend to full pulses and overall energy resolution. First, the aforementioned shift in the optimal \(\Delta t\) with varying signal strength \(1/\tau_{s}\) implies that as long as \(\Delta t\) is kept constant from measurement to measurement, there is not a single ``optimal'' \(\Delta t\) over a transient period of increased quasiparticle density. For example, for large energy depositions where \(\tau_{b}/\tau_{s}\) starts out very large, the optimum \(\Delta t\) may start lower than the selected \(\Delta t\), but then increase past the selected \(\Delta t\) as \(x_{qp}\) falls. As a result, tuning \(\Delta t\) may improve signal-to-noise in certain regions of a pulse but decrease signal-to-noise in other regions of the same pulse. Moreover, this trend implies that the dependence of the overall signal-to-noise (and hence energy resolution) on the search window duration itself has an overall energy dependence. While for conciseness we do not explore every detail of this and leave further exploration to future experimental work, we acknowledge that the overall dependence of energy resolution on the search window \(\Delta t\) is somewhat nontrivial.

\bibliography{main}{}

\end{document}